\documentclass[a4paper]{article}
\usepackage[left=1.4cm,right=1.4cm,top=1.9cm,bottom=1.4cm]{geometry}
\twocolumn
\usepackage[utf8]{inputenc}
\usepackage{graphicx}
\usepackage{amsmath}
\usepackage[super,comma]{natbib}
\usepackage{comment}
\usepackage{textcomp}
\usepackage{esint}
\usepackage[hyphens]{url}
\usepackage{xr}
\usepackage{authblk}
\usepackage{xcolor}
\usepackage{color}
\usepackage{comment}

\makeatletter
\newcommand*{\addFileDependency}[1]{
  \typeout{(#1)}
  \@addtofilelist{#1}
  \IfFileExists{#1}{}{\typeout{No file #1.}}
}
\makeatother

\newcommand*{\myexternaldocument}[1]{%
    \externaldocument{#1}%
    \addFileDependency{#1.tex}%
    \addFileDependency{#1.aux}%
}

\myexternaldocument{supplement}

\title{Unveiling the three-dimensional spin texture of skyrmion tubes}

\author[1,$\dagger$]{Daniel Wolf}
\author[2, 1,$\dagger$]{Sebastian Schneider}
\author[1]{Ulrich K. Rößler}
\author[3]{Andr\'as Kov\'acs}
\author[4]{Marcus Schmidt}
\author[3]{Rafal E. Dunin-Borkowski}
\author[1,5]{Bernd Büchner}
\author[2]{Bernd Rellinghaus}
\author[1,5,*]{Axel Lubk}

\affil[1]{Leibniz Institute for Solid State and Materials Research, IFW Dresden, Helmholtzstr. 20, 01069 Dresden, Germany}
\affil[2]{Dresden Center for Nanoanalysis, cfaed, Technische Universität Dresden, 01069 Dresden, Germany}
\affil[3]{Ernst Ruska-Centre for Microscopy and Spectroscopy with Electrons and Peter Grünberg Institute, Forschungszentrum Jülich, 52425 Jülich, Germany}
\affil[4]{Department Chemical Metal Science, Max Planck Institute for Chemical Physics of Solids,
Nöthnitzer Str. 40, 01187 Dresden, Germany}
\affil[5]{Institute of Solid State and Materials Physics, TU Dresden, 01069  Dresden,  Germany}
\affil[$\dagger$]{These authors contributed equally to this work.}
\affil[*]{Corresponding author.}

\begin{document}

\maketitle

\textbf{
Magnetic skyrmions \citep{Bogdanov1989} are stable topological solitons with complex non-co\-planar spin structures. Their nanoscopic size and the low electric currents required to initiate and control their motion has opened a new field of research, {\em skyrm\-ionics}, that aims at using skyrmions as information carriers for data storage and manipulation \citep{Kiselev2011,Sampaio2013,Tomasello2014,Fert2017}.}
\textbf{
Recent advances in skyrmionics prompt for a thorough understanding of the detailed three-dimen\-sional spin texture of a skyrmion \citep{Rybakov2013,Meynell2014,Leonov2016a,Schneider2018,Birch2020a,Yu2020} including skyrmion-skyrmion interactions and their coupling to surfaces and interfaces. These properties crucially affect application-related aspects such as the stability and mobility of skyrmions in confined structures.}
\textbf{
To date, however, experimental techniques to measure the three-dimensional (3D) spin texture with nanometer resolution are largely missing. We therefore adapt holographic vector field electron tomography\citep{Wolf2019} to the problem and report on the first quantitative reconstruction of the 3D spin texture of skyrmions with sub-10 nanometer resolution. The reconstructed textures reveal a variety of previously unseen local deviations from a homogeneous Bloch character within the skyrmion tubes (SkTs), details of the collapse of the skyrmion texture at surfaces, and a correlated modulation of the SkT in FeGe along their tube axes.
The quantitative 3D data of the magnetic induction also allow to experimentally confirm some principles of skyrmion formation by deriving spatially resolved maps of the magnetic energy density across these magnetic solitons.}

\section*{Introduction}

The unique features of magnetic skyrmions such as their competing magnetic interactions, topological structure, or dynamics are of great interest in both fundamental and applied physics. As multidimensional solitons, these particle-like states are localized in two-dimensions which requires a definite mechanism through additional frustrating magnetic couplings for their stabilization \citep{Roessler2006}. As consequence of their solitonic character, they can condense into thermodynamically stable phases, in particular dense packed lattices under applied fields \citep{Bogdanov1989}. The stabilization mechanism of these phases and their formation principles are ruled by effective skyrmion-skyrmion interactions \citep{Roessler2006}. However, the morphology of these phases in the phase-diagrams of real materials is dictated by the problem of condensation of two-dimensional periodic arrays, as in vortex-lattices of type-II superconductors \citep{Bogdanov1989}. In particular, the field-temperature phase diagram may hold various transitions between different condensed phases of skyrmions \citep{Roessler2011,Wilhelm2012}. Very recently, some studies address this problem for skyrmionic phases theoretically \citep{Balkind2019} and in experiments \citep{Huang2020}.  
In three dimensional bulk materials or thicker films the skyrmions are extended string-like objects; in the simplest formation they are homogeneously continued as skyrmion tubes (SkT) preserving translational invariance along their axis. In magnetic nanoobjects, however, the influence of surfaces will affect formation, shape and interaction of skyrmions and the stabilization of condensed skyrmionic phases. Already from the earliest observations of skyrmionic phases in films of chiral helimagnets \citep{Yu2011,Wilson2012}, it is known that their phase diagram massively deviate form those of bulk materials. It has been shown that 3D surface twists can stabilize SkTs in thin films \citep{Rybakov2013,Meynell2014,Rybakov2016,Leonov2016} and that 3D modulations of SkTs embedded in a conical host phase may introduce an attractive interaction between these tubes \citep{Du2018}. 3D SkT modulations also affect emergent electric and magnetic fields acting on spin-polarized electrons and magnons \citep{Nagaosa2013}, which results in unusual transport phenomena \citep{Leonov2017} on top of the "normal" topological Hall effect in static and current-driven skyrmion crystals \citep{Lee2009,Neubauer2009,Zang2011,Schulz2012}.

\begin{figure*}[ht]
    \centering
    \includegraphics[width=0.9\textwidth]{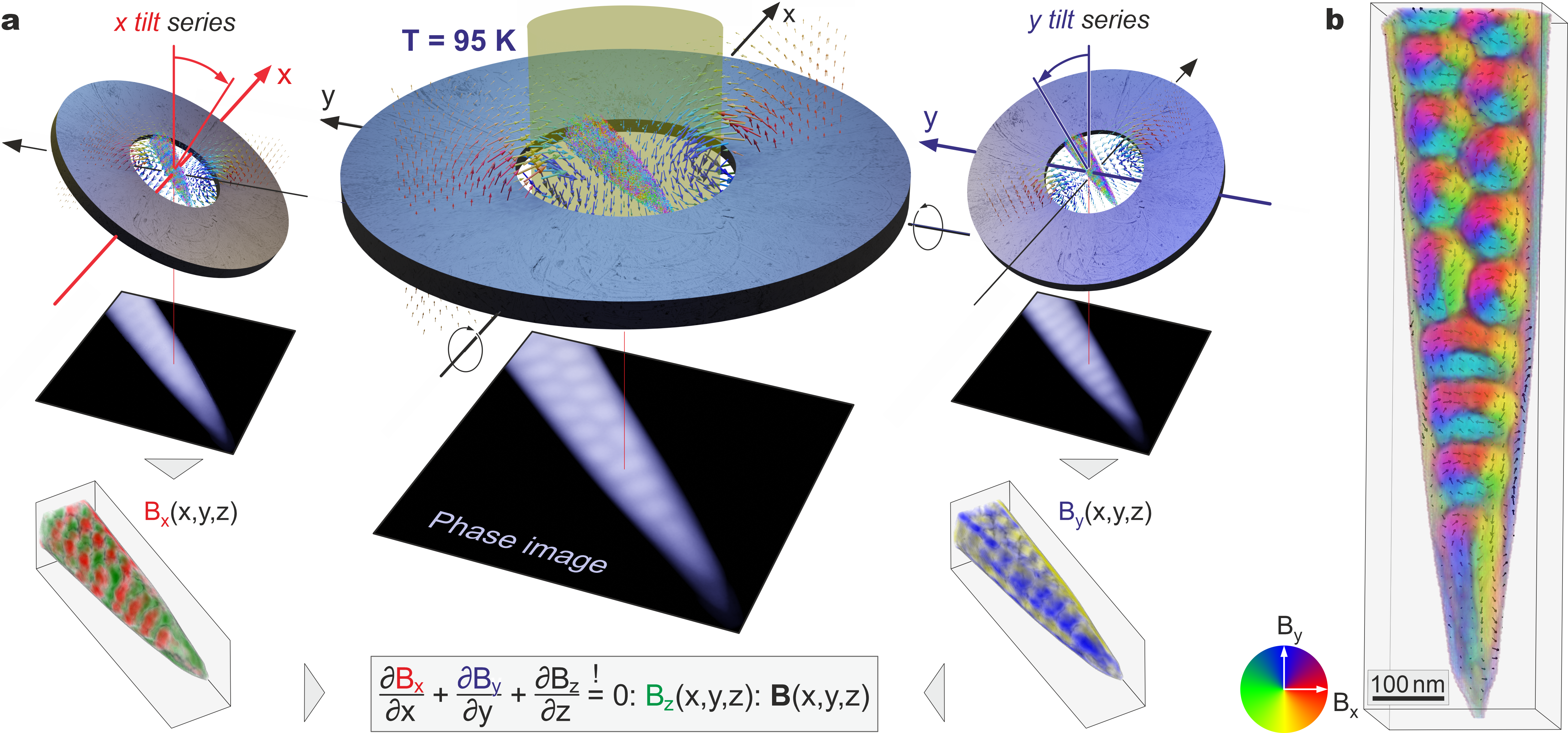}
    \caption{\textbf{a}, Holographic vector-field electron tomography (VFET) of skyrmions in FeGe. A needle-shaped sample is placed above an out-of-plane magnetized $\mathrm{Sm_2Co_{17}}$ ring in a liquid nitrogen cooled TEM holder. The low temperature and the remanent stray field of the ring stabilizes skyrmion tubes and their orientation with respect to the holder. A tilt series of 2D phase images of the transmitted electron wave is obtained from off-axis electron holographic tilt series around the $x$ (left) and $y$ axes (right). Subsequently, $x$ and $y$ components of the magnetic induction $\boldsymbol{B}$ are tomographically reconstructed from the corresponding phase tilt series. Solving $\mathrm{div}\boldsymbol{B}=0$ finally yields the $z$ component, hence the full 3D vector of $\boldsymbol{B}(x,y,z)$. \textbf{b}, 3D map of the resulting magnetic induction. For clarity, only the experimental $B_x$ and $B_y$ components are shown here.}
    \label{main:Fig_1}
\end{figure*}

Si\-mi\-larly, e.g., in hybrid chiral ferromagnet/super\-con\-ductor systems \citep{Dahir2019}, the functionalization of skyrmions is modified by 3D modulations of the SkT at the interface. Finally, the observation of unusually strong topological quantum Hall effects \citep{Huang2012} may indicate the presence of abrupt magnetization changes such as in magnetic bobbers (incorporating a Bloch point) in surface regions \citep{Zheng2018}.

Notwithstanding the importance of 3D effects, neither exact 3D models of SkTs in realistic confined geometries nor high-resolution experimental mappings of their spin texture are currently available, although effects of confinement \citep{Wilson2012, Rybakov2013,Rybakov2016} and of anisotropies \citep{Leonov2020} have received attention.  This lack of data prevents a deeper understanding of skyrmion lattice defects \citep{Yu2020,Jin2017}, influence of surface anisotropies, curvatures \citep{Kravchuk2018}, and real structure effects in the modulation of 3D skyrmionic spin textures.
Among the various high-resolution magnetic imaging techniques, transmission electron microscopy (TEM) based electron holography (EH)\citep{Phatak2010,Tanigaki2015,Wolf2015,Wolf2019} and X-ray magnetic chiral dichrosim (XMCD) \citep{Streubel2015,Donnelly2017,Hierro-Rodriguez2020} can be conducted in a tomographic way to determine the 3D magnetic induction, $\boldsymbol{B}$, or magnetization, $\boldsymbol{M}$, of a sample, respectively. In this work, we employ holographic vector-field electron tomography (VFET) as it provides a higher spatial resolution (below 10 nm) than X-ray based methods, which is crucial for resolving the details of magnetic textures in nanomagnetic structures such as vortices \citep{Wolf2019} or skyrmions. The limited space in a high-resolution TEM instrument, however, has so far prevented any in-situ applications of rotatable (out-of-plane) magnetic fields to a cryogenically cooled sample, which is essential for the acquisition of tomographic tilt series of electron holograms from a sample that needs to be magnetically stabilized. This limitation impedes the measurement and 3D reconstruction of spin textures for a large class of materials with a metastable skyrmion phase at non-zero applied fields below room temperature (e.g., many isotropic helimagnets). For the present experiments, we have therefore devised a setup that overcomes these obstacles.

\section*{High resolution vector-field tomography in an external magnetic field} 

\begin{figure*}[t]
    \centering
    \includegraphics[width=0.95\textwidth]{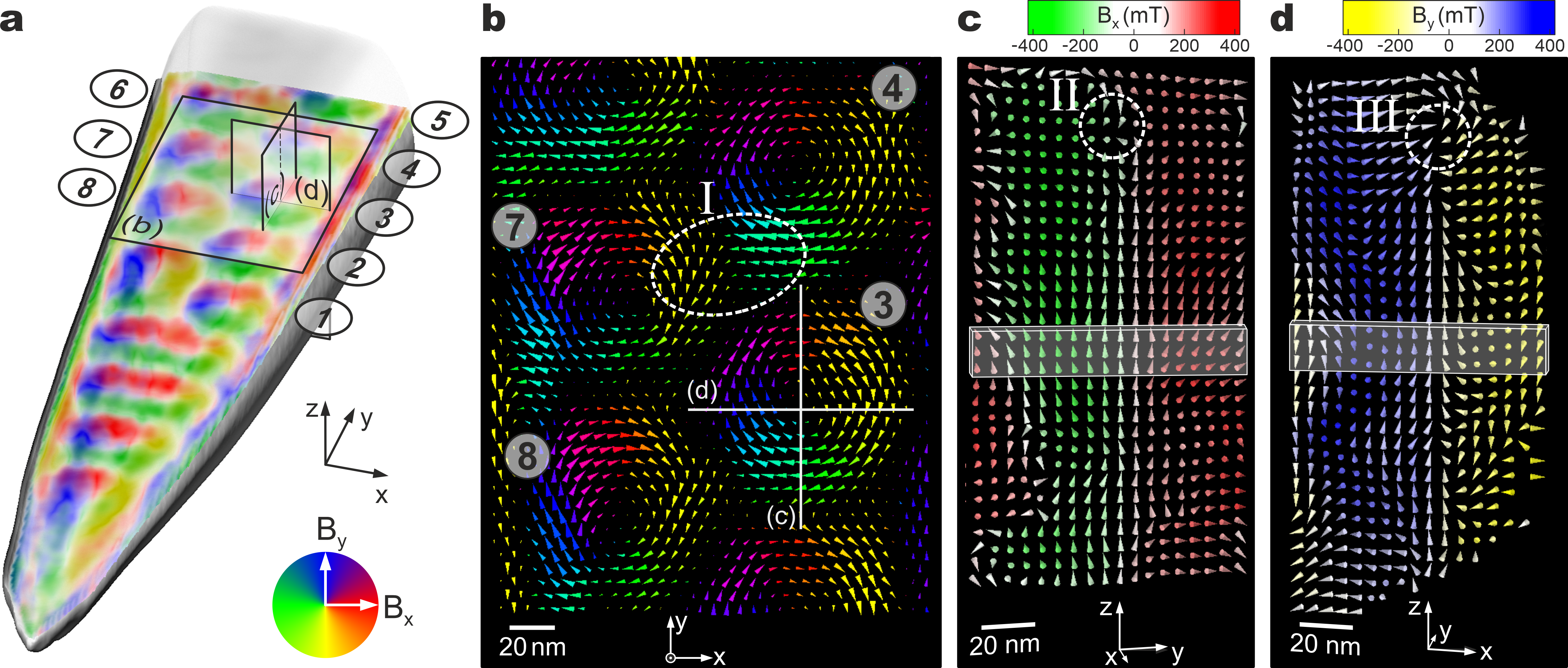}
    \caption{Spin texture of skyrmion tubes in a FeGe needle: \textbf{a}, Volume rendering (colored) of the in-plane ($x$, $y$) components of the reconstructed magnetic induction $\boldsymbol{B}$ and iso-surface of the mean inner potential highlighting the sample shape (grey, bottom half only). Rectangles indicate cross-sections, whose details are shown in \textbf{c} and \textbf{d}. \textbf{b}, Planar $x$-$y$ cross-section. Color and size of the arrows indicate the direction and magnitude of the in-plane component of $\boldsymbol{B}$, respectively. \textbf{c} and \textbf{d}, direction of $\boldsymbol{B}$ (arrow orientation) and magnitude (color) of $B_x$ and $B_y$ in $y$-$z$ and $x$-$z$ cross-sections through SkT 3. Here, the SkT was artificially aligned along its $z$ axis (see text for details).}
    \label{main:Fig_2}
\end{figure*}

The tomographic investigation of the magnetic texture of skyrmions was conducted on a sample of the isotropic helimagnet FeGe with $\mathrm{P2_13}$ structure (B20 phase). The material was chosen, since FeGe is an otherwise well studied archetypical skyrmion host with a rather large skyrmion phase pocket in the phase diagram spanned by temperature and external field \citep{Yu2011,Stolt2019}. A needle-shaped sample (cf. Supplementary Sect.~\ref{suppl:note:geometry}) was cut from a FeGe single crystal by focused ion beam (FIB) including ion polishing to restrict the ion beam damage to a surface layer of some nanometers (see Methods). The dimensions and shape of the needle ensure that, even at high tilt angles, the sample is fully electron-transparent and the obtained holographic projections cover the same sample region. Additionally, the elongated shape has some technological significance for anticipated spintronic devices such as racetrack memories \citep{Parkin2008}. In order to (i) adjust the skyrmion phase below the Curie temperature and (ii) stabilize the orientation of the skyrmion lattice with respect to the TEM holder, the FeGe needle was steadily exposed to an out-of-plane magnetic field of $\mu_0H_\mathrm{ext} \approx 170\, \mathrm{mT}$. The field was provided through the remanent stray field of a ring-shaped $\rm Sm_2Co_{17}$ hard magnet that was placed under the sample in a tomography-adapted liquid nitrogen TEM cooling holder (cf.\ Fig.~\ref{main:Fig_1}a). The field is virtually homogeneous across the micron-sized sample (cf.\ Supplementary Sect.~\ref{suppl:note:ring}).

Using this special setup, we have recorded three holographic tilt series as required for VFET \citep{Tanigaki2015,Wolf2019} (see Methods and Supplementary Sect.~\ref{suppl:note:geometry} for details of the imaging conditions). The first series of holograms was acquired by tilting the sample around the $x$ axis at room temperature, since above the Curie temperature of $\rm T_C = 278.7\,K$ \citep{Kovacs2017}, the phase $\varphi_e$ reconstructed from the holograms is of pure electrostatic origin. The (scalar) electrostatic potential $\rm \Phi$ was then determined by inverting the Radon transformation (i.e., linear projection law) linking $\rm \Phi$ and $\varphi_e$. The resulting 3D mean inner potential distribution is nearly homogeneous as discussed in Supplementary Sect.~\ref{suppl:note:MIP} (see Methods for the tomographic reconstruction details). 

In the following two series, the sample was tilted around the $x$ and $y$ axes (cf.\ Fig.~\ref{main:Fig_1}a) at $T = 95\,\mathrm{K}$. At this temperature below $T_C$, the magnetic fields imposes an additional Aharonov-Bohm phase $\varphi_m$ on the imaging electrons. After subtracting the pre-determined electrostatic contribution from the total phase shift, the in-plane components of the magnetic induction, $B_x(x,y,z)$ and $B_y(x,y,z)$ (see Fig.~\ref{main:Fig_1}b), were reconstructed from the remaining $\varphi_m$ in 3D by inverse Radon transformation of another linear projection law linking the gradient of $\varphi_m$ and $B_{x,y}$ (see Methods for details). The spatial resolution of the reconstructed $B_x$ and $B_y$ components was below $10\, \mathrm{nm}$ in directions outside of the missing tilt range (cf.\ Supplementary Sect.~\ref{suppl:note:resolution}).

In order to change the tilt axis from $x$ to $y$, the sample required to be warmed up to room temperature and rotated in-plane by $90^\circ$ in the sample holder. The associated heating and (field) cooling of the sample followed precisely the same protocol as prior to tilting it around $x$. As a result, at the here investigated and most confined tip region of the FeGe needle, the skyrmion patterns obtained after cooling prior to acquiring the $x$ and $y$ tilt series were almost perfectly identical, while at the less confined broader end of the needle, the skyrmion pattern had changed considerably (see Suppl.\ Fig.~\ref{suppl:Fig_Specimen} for details).

Based on $B_{x,y}$, also the remaining third component $B_z(x,y,z)$ was finally determined by solving $\mathrm{div}\,\boldsymbol{B}=0$, thereby yielding the full 3D vector field of the magnetic induction $\boldsymbol{B}(x,y,z)$ (see Methods for details and Supplementary Movie 1 for 3D animations of the tomograms). Since the calculation of $B_z$ is, however, based on the differentiation of $B_x$ and $B_y$, it suffers from noise amplification and corresponding artefacts, which needs to be taken into account in the following analysis.

\section*{Spin texture of Skyrmion tubes} 

In the following, we analyze this comprehensive 3D set of $\boldsymbol{B}(x,y,z)$ data in order to extract characteristic magnetic features and quantities of the SkTs in FeGe. Fig.~\ref{main:Fig_1}b reveals that the tip of the needle hosts a single row of SkTs that are elliptically distorted towards the sideward surfaces of the needle, i.e., perpendicular to both the needle axis and the stabilizing external field. Towards the broader back of the needle (top region of Fig.~\ref{main:Fig_1}b), these elongated SkTs develop into a zig-zag chain of Bloch SkTs, when the width surpasses a critical value of roughly 150\,nm. This width corresponds to about twice the characteristic helical modulation length $L_D$ and the next-nearest neighbour distance in a close-packed skyrmion lattice in FeGe \citep{Jin2017}. An evaluation of the out-of-plane component of $\boldsymbol{B}$ (cf.\ Supplementary Sect.~\ref{suppl:note:Lattice Packing}) reveals a ratio of areas with positive and negative $B_z$ of 0.85, which also points to a close-packing of the SkTs in this region \citep{Balkind2019}.

\begin{figure}[t]
    \centering
    \includegraphics[width=0.9\columnwidth]{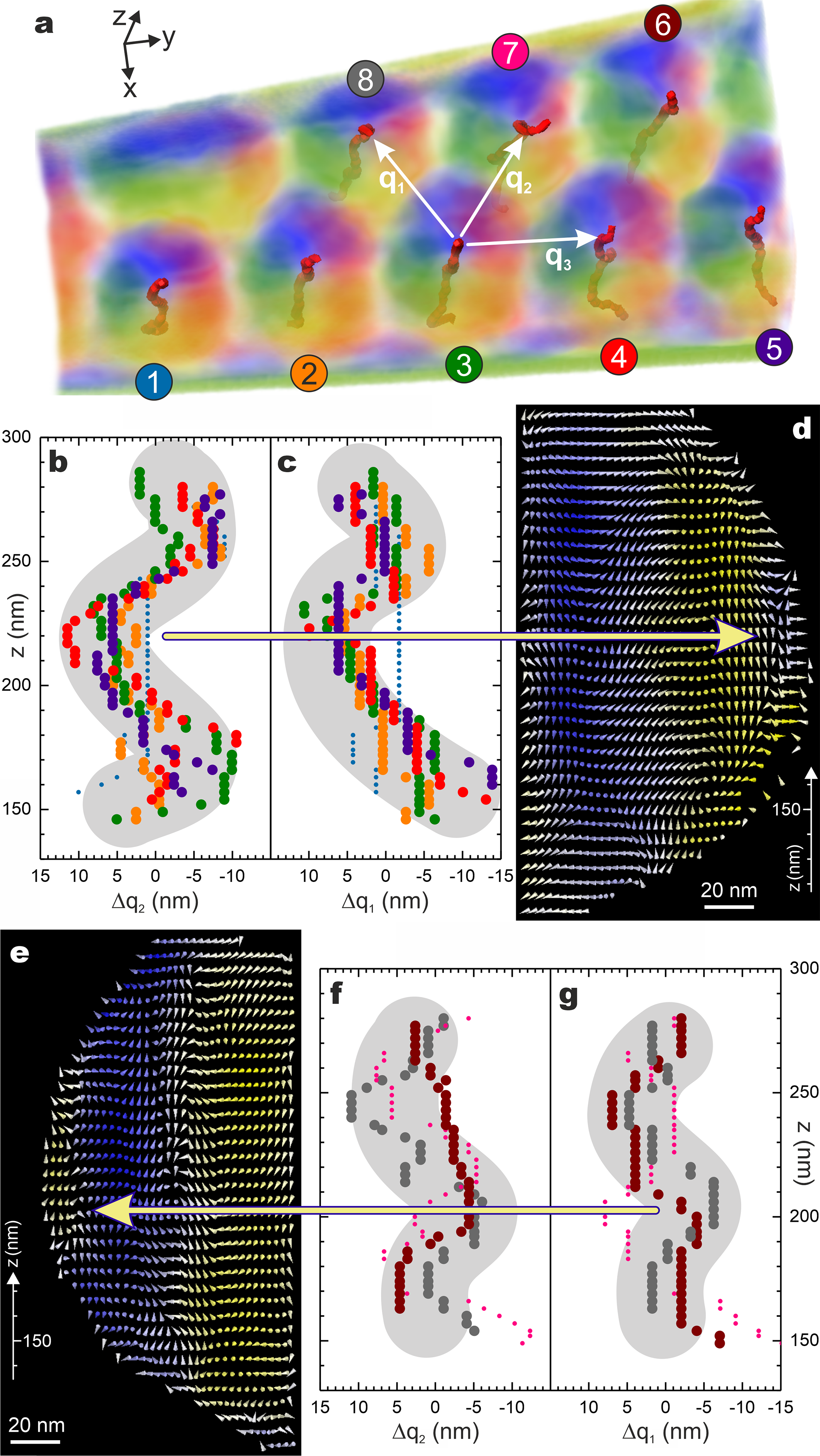}
    \caption{Axial modulations of the SkTs. \textbf{a}, 3D rendering of the ($x$, $y$) components of $\boldsymbol{B}$ for eight close-packed SkTs in the FeGe needle. The red lines represent the SkT axes, and $\boldsymbol{q_1}$, $\boldsymbol{q_2}$, $\boldsymbol{q_3}$ indicate the three NN directions of the SkT lattice. \textbf{b} and \textbf{c}, $z$ dependent positions of the cores of SkTs 1-5 and SkTs 6-8 (\textbf{f} and \textbf{g}) along $\boldsymbol{q_2}$ and $\boldsymbol{q_1}$. The data points are colored according to the labels in \textbf{a}. \textbf{d} and \textbf{e}, $x$-$z$ slices through SkTs 3 \textbf{d} and 8 \textbf{e}.}
    \label{main:Fig_3}
\end{figure}

%
%
Fig.~\ref{main:Fig_2} and Supplementary Movies 2, 3 represent an in-depth view of the spin texture within these SkTs. 
Fig.\ \ref{main:Fig_2}b shows a planar cross-section of the spin texture through the vertical center of the needle.
Here, the color of the arrows indicate the direction of the in-plane ($x$, $y$) components of $\boldsymbol{B}$ according to the color wheel in Fig.\ \ref{main:Fig_2}a. 
While all four SkTs in this section feature radial Bloch walls, details of the spin texture exhibit subtle discrepancies from that of an undisturbed, perfect Bloch skyrmion: (i) The skyrmions may exhibit significant distortions and (partially) lose their axial symmetry. Neither the direction of the in-plane component of $\boldsymbol{B}$ remain tangential nor is its magnitude constant for a given radius. (ii) Frequently, the maximum out-of-plane orientation (as indicated by a virtually vanishing size of the arrows) is not located in the center of the skyrmions. (iii) Unlike expected for isolated magnetic solitons, some distortions of the skyrmionic spin textures seem to go along with magnetic flux "leaking" between neighboring SkTs as highlighted by the dashed region "I" between SkTs 4 and 7. The mere similarity of this appearance with a confined helical "band" resembles the evolution of a metastable isolated skyrmion towards a helical modulation ("strip-out") discussed in bi-layer thin films \citep{Leonov2016b}. Both scenarios, however, violate the topology of a skyrmion and necessitate the occurrence of singular Bloch points.
Features indicating the occurrence of Bloch points are in fact frequently observed: Figs.\ \ref{main:Fig_2}c, d show two orthogonal vertical slices through SkT 3. Since the SkT axes are found to be bent and (partially) twisted (see below and Fig.\ \ref{main:Fig_3}), SkT 3 was artificially aligned along the $z$ axis for this presentation.
To this end, each $x$-$y$ slice of the tube was laterally shifted such that the minima of the in-plane component $B_{\|}=\sqrt{B_x^2+B_y^2}$ of all slices are aligned along the $z$ axis. The dashed circles labeled "II" and "III" highlight sections that are indeed reminiscent of a Bloch point. Besides, both cross-sections confirm the lack of axial symmetry and substantiate the overall inhomogeneity of the spin texture in the SkT already seen from the planar cross-section. In contrast to "pure" Bloch SkT, we frequently (but not systematically) observe a (partial) radial (i.e., a small N\'{e}el character) orientation of the local magnetic induction. These imperfections grow upon approaching the surface and finally lead to a total collapse of the skyrmion structure. This becomes most apparent in the $x$-$z$ cross-section in Fig.\ \ref{main:Fig_2}d, where the thickness of the needle decreases. This region should be understood as result of surface symmetry breaking and concomitant  effects, such as pinning by surface anisotropies, modified magnetic properties due to FIB surface damage and demagnetization fields.

\begin{figure}[t]
    \centering
    \includegraphics[width=7cm]{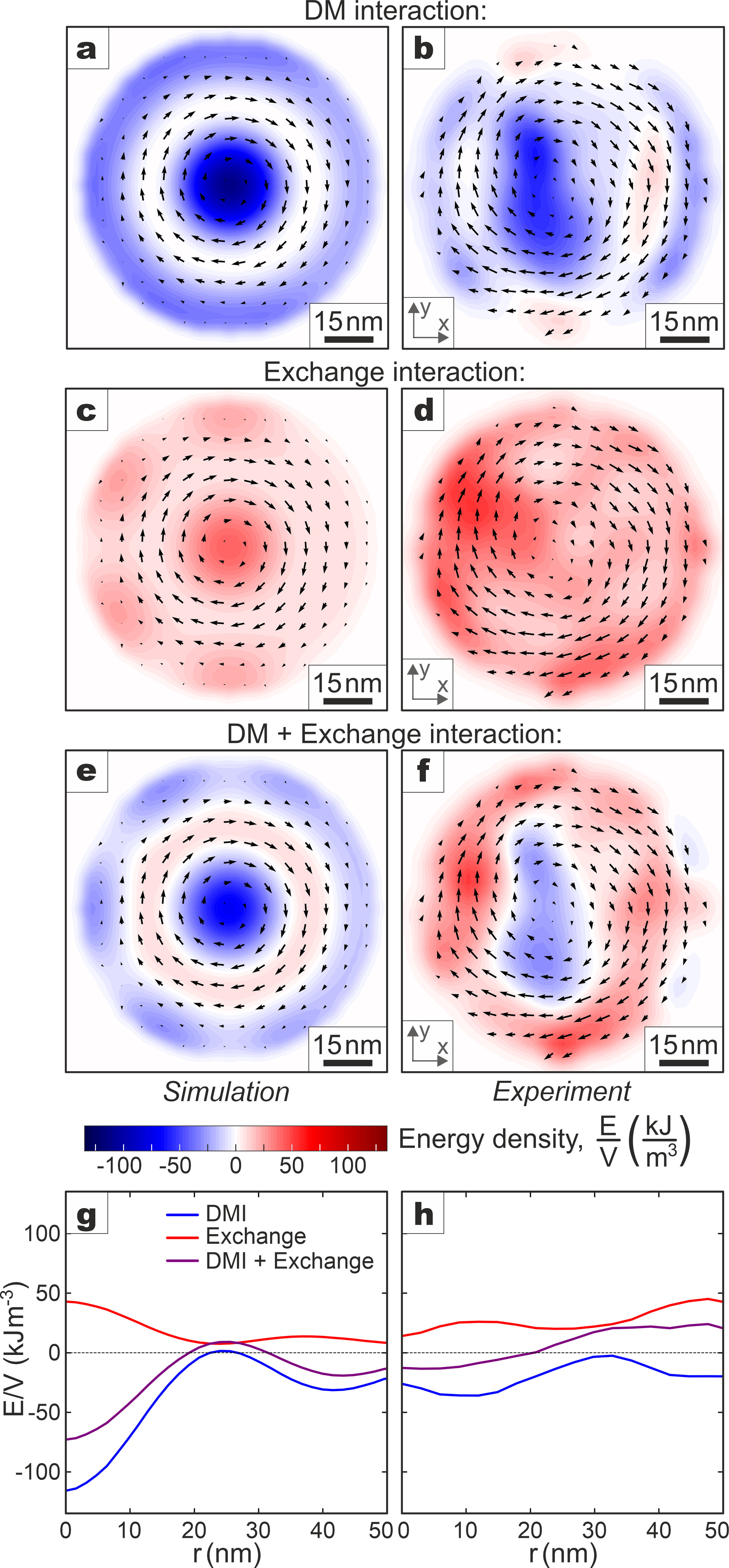}
    \caption{Planar maps of the predominant magnetic energy density contributions $DB_z ( \nabla \times \boldsymbol{B})_z$ (DM interaction, (\textbf{a} and \textbf{b})) and $A\|\left(\nabla \times \boldsymbol{B}\right)_z \|^2$ (exchange interaction, (\textbf{c} and \textbf{d})) and their sum (\textbf{e} and \textbf{f}), together
    with radial averages as function of the radius (\textbf{g}, \textbf{h}). Left: Simulations. Right: Reconstructed from the central part of SkT 3. Overlaid arrows indicate the in-plane magnetic induction.}
    \label{main:Fig_4}
\end{figure}

In the cubic helimagnet FeGe, a twisting in the third direction, i.e. along the axis of the SkT, could result in a gain of energy through the Dzyaloshinskii-Moryia (DM) exchange. However, such an effect will not create triply twisted structures of skyrmions, as this is geometrically impossible, because the ferromagnetic vector can be rotated only in two directions in the cutting plane perpendicular to the skyrmion axis. Hence, the chiral twist\citep{Rybakov2013} could only affect the shape of the SkTs. E.g., a modulation could arise as a  tertiary conformational  deviation  from straight cylindrical SkT shape. Confinement or surface pinning may promote such morphology changes. Indeed, Fig.\ \ref{main:Fig_3}a illustrates that the axis of SkTs (red lines) are axially bent and twisted rather than extending as straight cylindrical objects along the $z$ axis in the close-packed region of the needle (similar rendering of $B_{\|}$ as in Fig.\ \ref{main:Fig_2}a). In order to study possible correlations between these deformations, we have analyzed the in-plane positions of the SkT axes along the nearest neighbour (NN) directions $\boldsymbol{q_1}$, $\boldsymbol{q_2}$, $\boldsymbol{q_3}$ (indicated by white arrows). The resulting dependencies of the deviations from an average axial position along $\boldsymbol{q_2}$ and $\boldsymbol{q_1}$, i.e., in directions that are largely affected by the lateral confinement, are shown in Fig.\ \ref{main:Fig_3}b, c for the bottom row of SkTs (nos.\ 1-5) and in Fig.\ \ref{main:Fig_3}f, g for SkTs 6-8 in the top row. Except for SkT 1 (small blue circles in b and c), which is least close-packed and rather has two elliptically elongated SkT neighbours, and SkT 7 (small pink circles in b and c), which is additionally distorted due to an unusual magnetic coupling to SkT 4 (see above), all SkT axes exhibit pronounced sideward deformations. As indicated by grey bands (guides to the eye only), these lateral modulations are correlated among SkTs in the same row. These deformations are harmonically modulated with a modulation length of approximately $\rm 80\,nm$ that is close to the helical period $L_D \simeq 70\, \mathrm{nm}$ in FeGe \citep{Yu2011,Lebech1989} pointing to the DM interaction as a possible origin of the deformations. Note, however, that comparisons with $y$-$z$ cross-sections through SkTs 3 and 8 in Figs.\ \ref{main:Fig_3}d, e reveal that these modulations correlate with the occurrence of uniformly polarized edge states \citep{Song2018}. These edge states reside at the sidelong rims of the FeGe needle (cf.\ left and right surfaces in Figs.\ \ref{main:Fig_1}b and \ref{main:Fig_2}a) and are separated by a very narrow magnetic transition regions (resembling domain walls) of some 10\,nm in width from the SkTs. The correlation of the deformation of the SkTs with these edge states is corroborated by the facts that (i) the central deformations are directed towards the center on both sides of the needle and (ii) the magnetic orientations of the edge states and the outer rims of the SkTs' spin textures are concurrently reversed between the right (SkTs 1-5) and left side (SkTs 6-8) of the needle, respectively. This results in qualitatively identical interactions between the SkTs and the edge states on either side. In contrast, the deformations of the SkTs along the largely unconfined direction $\boldsymbol{q_3}$ do {\em not} exhibit any obvious correlations (not shown).

The availability of 3D vector data of the magnetic induction enables us for the first time to experimentally derive from the volume of a sample spatial maps of free energy density contributions from magnetic exchange and DM interactions, respectively. These energetic contributions are most essential for the formation and stabilization of skyrmions and SkTs, as they are expected to reduce the free energy in the centers of the SkTs, while the interstitial regions in a SkT lattice may be considered as domain walls of increased energy \citep{Butenko2010}. 
We have calculated from $\boldsymbol{B}(x,y,z)$ the solenoidal part of the magnetic exchange energy density $\mathit{w}_\mathrm{ex}=A\|(\nabla \times \boldsymbol{B}) \|^2$ and the volume contribution of the DM energy density $\mathit{w}_\mathrm{DM} = D \boldsymbol{B} \cdot ( \nabla \times \boldsymbol{B})$.
Here, $A = 8.78 \ \mathrm{\frac{pJ}{m}}$ and $D = 1.58 \ \mathrm{\frac{mJ}{m^2}}$ denote the exchange stiffness and the DM interaction strength of FeGe, respectively \citep{Beg2015}. Due to the vanishing magnetic charge density $\rho_m \approx 0$, the conservative part of the exchange energy $\|\nabla \cdot \boldsymbol{M}\|^2$ is small in Bloch skyrmions, and other contributions are divergence contributions that can be collapsed to surface terms, which is the reason why both are neglected here. As the spin texture of the SkTs is highly disturbed in the near-surface region (cf.\ Fig.\ \ref{main:Fig_2}c, d), and in order to account for the axial deformation of the SkTs, only magnetic induction data from the central part of the SkT (cf.\ grey shaded boxes in Figs.\ \ref{main:Fig_2}c, d) was used and projected in the $x$-$y$ plane to calculate the planar distribution of energy densities. For comparison, such energy density maps were also calculated for a simplified skyrmion lattice model employing the circular cell approximation \citep{Bogdanov1994} taking into account the shape of the needle (Supplementary Sect.~\ref{suppl:note:Magnetostatics}). 

Fig.\ \ref{main:Fig_4} shows the resulting simulated (left column) and experimentally determined energy density maps (right column) for the contributions arising from the DM and exchange interactions, and their sum. Here, we only plot energy densities dominated by the in-plane components of the magnetic induction to suppress some of the artefacts afflicting the $B_z$ component, which is, however, sufficient and consistent with calculations that take the full $\boldsymbol{B}(x,y,z)$ into account (cf.\ Supplementary Sect.~\ref{suppl:note:mag-energies} and Fig.\ \ref{Suppl_Fig_energetics_simulation}). Given the apparent real structure effects and noise in the data, the experimental results are in excellent qualitative agreement with the simulations. In particular, the course of the radially averaged contribution (g, h) confirms experimentally the prediction that the reduction in the free energy density due to the DM interaction overcompensates the energetic costs of the exchange in the core of the skyrmion tube, which results in the overall energetic stabilization of the SkT (lattice). In comparison with the simulations, the experimental energy density landscapes are slightly compressed in $x$ direction, which is attributed to the interaction with the edge state and contributes, besides the noise, to the quantitative reduction of the radial averages in Fig.\ \ref{main:Fig_4}h. Noteworthy, the energetically least stable part of the SkT is the center of the circumferential Bloch wall, which is the region with the highest in-plane orientation of the magnetic induction in the SkT. It is remarkable that this is precisely that region, where we had found indications of "leaking flux" between SkTs 4 and 7 (see discussion above and Fig.\ \ref{main:Fig_2}b). Apparently, the center of the Bloch wall is the most "forgiving" zone of the SkT, which in turn may explain its overall stability against the variety of observed magnetic defects. All in all the energy maps confirm the heterogeneous particle-like nature of skyrmions. Unlike the energetically homogeneous spiral states of a chiral helimagnet like FeGe, the skyrmions have a definite shape and size that is caused by the frustration between the different exchange energies, which can be lifted only partially through the doubly twisted core.

\section*{Summary}

Low temperature holographic vector-field electron tomography was used in combination with the spatial stabilization of the specimen's magnetic state by an external magnetic field to reconstruct the full vector-field $\boldsymbol{B}$ of the skyrm\-ionic spin texture in FeGe in all three dimensions at nanometer resolution. The unrivaled resolution of this 3D magnetic microscopy of a volume sample reveals unprecedented insight into the details of the 3D spin texture of skyrmions.

Besides a characterization of the complicated breakdown of the skyrmion texture upon approaching surfaces in axial directions, we observe a variety of imperfections in the spatial extension of skyrmion tubes. Among them are axial and planar distortions of the SkTs, local losses of axial symmetry, and the occurrence of unexpected radial (rather than purely tangential) tilts of the magnetic induction in the circumferential Bloch walls. Even indications of in-plane magnetic flux leaking among neighboring SkTs in close-packed regions and abrupt changes of the magnetic induction that may be indicative of the occurrence of Bloch points are found. Also, the 3D course of the SkT axes was investigated in great detail. Here, we observe a substantial bending and twisting of these axes that is locally correlated with the occurrence of pronounced edge states, specifically in directions that are affected by confinements. Noteworthy, these deformations appear at length scales, where harmonic modulations are promoted by the DM interaction.

Planar energy density maps across the SkTs were derived from the volume data of the magnetic induction and confirm for the first time experimentally the anticipated formation and stabilization mechanism of skyrmions for a volume sample. The results reveal a substantial energetic gain due to the DM interaction that overcompensates the energetic effort associated with the magnetic exchange interaction in the core of the SkT thereby stabilizing the SkT lattice as a whole.

We anticipate that this novel experimental approach will pave the way to a better understanding of spin textures in a large variety of complex topologically protected and non-topological magnetization patterns, including other members of the skyrmion family, thereby moving the fields of both nanomagnetism and spintronics significantly forward.

\section*{Methods}
\subsection*{Sample preparation.}

Based on the results of crystal growth by chemical vapour transport in the system Fe/Ge \citep{Richardson1967,Bosholm2001} single crystals of FeGe in the B20 structure were grown via chemical transport reaction using iodine as transport agent.
Starting from a homogeneous mixture of the element powders iron (Alfa Aesar 99,995\,\%) and germanium (Alfa Aesar 99,999\,\%) the cubic modification of FeGe crystallized by a chemical transport reaction very slowly in a temperature gradient from 850 K (source) to 810 K (sink), and a transport agent concentration of $0.2 \, \mathrm{\frac{mg}{cm^3}}$ iodine (Alfa Aesar 99,998\,\%). The chemical vapour transport was made perpendicular to the tube axis over a diffusion distance of 38\,mm. Selected crystals were characterized by EDXS, WDXS, and especially X-ray single crystal diffraction to verify the present modification.

The preparation of the FeGe needle was carried out via focused ion beam (FIB) technique on a Thermo Scientific Helios 660. A rough cut of the needle geometry $(700 \times 700 \, \mathrm{nm})$ was performed with currents of 790 and $430 \, \mathrm{pA}$ respectively. For further fine shaping $(300 \times 300 \, \mathrm{nm})$ the current was reduced to 80 and $40 \, \mathrm{pA}$. The final polishing was carried out at $24 \, \mathrm{pA}$. In order to remove preparation residue the needle was finally cleaned in a Fischione Model 1070 NanoClean for $1 \, \mathrm{min}$.  

High-resolution TEM images indicate, that the crystalline core of the needle is surrounded by an amorphous surface layer of 4\,nm. STEM-EDX measurements reveal, that this layer consists of iron oxide (cf. Supplementary Sect.~\ref{suppl:note:surface layer}).

The skyrmion phase within the needle was stabilized by the stray field of a ring-shaped $\mathrm{Sm_2Co_{17}}$ permanent magnet fitted into a GATAN 636 double tilt liquid nitrogen sample holder. The ring was prepared from a bulk magnet by sinker spark eroding and mechanical grinding and provided a magnetic field in $z$-direction of approximately $170 \, \mathrm{mT}$ (cf.\ Supplementary Sect.~\ref{suppl:note:ring}).

\subsection*{Acquisition and reconstruction of the holographic tilt series.} 

Holographic tilt series were recorded at an FEI Titan G2 60-300 HOLO\citep{Boothroyd2016} in Lorentz-Mode (conventional objective lens switched off) operated at $300\,\mathrm{kV}$. The voltage of the electrostatic M\"ollenstedt bisprism was set to 120 V leading to a fringe spacing of $2.3\,\mathrm{nm}$ in the electron hologram (cf.\ Supplementary Sect.~\ref{suppl:note:solitonic skyrmions}). For the acquisition of the latter, a GATAN K2 Summit direct detection camera in counting modes was used yielding a holographic fringe contrast of $40\,\%$. The acquisition process was performed semi-automatically with an in-house developed software package \citep{Wolf2010} to collect three holographic tilt series consisting of object and object-free empty holograms, two at $95\,\mathrm K$ and one at room temperature. For the first tilt series at $95\,\mathrm K$, the angle between the needle and tilt axis amounts to $30^{\circ}$. For the second tilt series, the specimen was manually rotated outside the microscope in-plane by $70^{\circ}$ (ideal is $90^{\circ}$) resulting in an angle between the needle and tilt axis of $-40^{\circ}$ (cf. Supplementary Sect.~\ref{suppl:note:geometry} for the details). The tilt range of each tilt series was from $-66^{\circ}$ to $+65^{\circ}$ in $3^{\circ}$ steps. To obtain the full phase shift ($>2\pi$), the phase images were unwrapped automatically by the Flynn algorithm \citep{Ghiglia1998} and manually at regions, where the phase signal is too noisy or undersampled, by using prior knowledge of the phase shift (e.g., from adjacent projections) \citep{Lubk2014}. Potential phase wedges in vacuum caused by the magnetic stray-field of the ring were corrected in all three tilt series. An analysis of these stray-field contributions is presented in the Supplementary Sect.~\ref{suppl:note:el_mag_fringing_fields}.

\subsection*{Tomographic reconstruction.} 
All three phase image tilt series were aligned, i.e., corrected for image displacements with respect to their common tilt axis by cross-correlation, center-of-mass method and common-line approach\citep{Wolf2019}. Thereby obtained aligned datasets correspond to the following linear projection laws (Radon transformations): 
\begin{equation}
    \varphi_e\left(p,\theta,z\right)=C_E\iint_{\boldsymbol{e}\cdot\boldsymbol{r}}\Phi\left(x,y,z\right) \mathrm{d}x\mathrm{d}y
\end{equation}
and
\begin{equation}
\frac{\partial\varphi_m(p,\theta,z)}{\partial z}=\frac{e}{\hbar}\iint_{\boldsymbol{e}\cdot\boldsymbol{r}} B_{p=y,x}\left(x,y,z\right) \mathrm{d}x\mathrm{d}y.    
\end{equation}
Here, $C_E$ is a kinetic constant depending solely on the acceleration voltage, $p$ and $z$ are the 2D detector coordinates, $\theta$ the tilt angle, $\boldsymbol{r}=(x,y)^T$, and $\boldsymbol{e} = (\cos{\theta},\sin{\theta})^T$. The index to the integral indicates a collapse of the 2D integral to the projection line defined by $\boldsymbol{e}\cdot\boldsymbol{r}$. The subsequent tomographic 3D reconstruction of the aligned phase tilt series (i.e., the inverse Radon transformation) was numerically carried out using weighted simultaneous iterative reconstruction technique (W-SIRT) \citep{Wolf2014}.

The three resulting tomograms represent the incremental 3D phase shift per voxel that we refer to as 3D phase maps. The two 3D phase maps obtained at $95\,\mathrm K$ were released from their electrostatic (MIP) contribution by superposition and subtraction of the 3D phase map obtained at room temperature. Then, the derivation of each of the two resulting magnetic 3D phase maps in direction perpendicular to the experimental tilt axis and multiplication with the factor $\hbar /e$ leads to one component of the magnetic induction in the respective direction. Since the specimen was rotated only by $70^{\circ}$ in the underlying tomographic experiment for the reconstruction of these two $\boldsymbol{B}$-field components, one of them was projected on the orthogonal direction of the other to receive finally the 3D $B_x$ and $B_y$ component. A verification of the experimental workflow repeated on simulated data is provided in Supplementary Sect.~\ref{suppl:note:Rec-of-Sim}.

\subsection*{Calculation of the third magnetic B field component.}
The third $\boldsymbol{B}$ field component $B_z$ is obtained by solving Gauss's law for magnetism $\mathrm{div}\,\boldsymbol{B}=0$ with appropriate boundary conditions on the surface of the reconstruction volume. Here, we employed periodic boundary conditions for solving this differential equation in Fourier space endowed with coordinates $\boldsymbol{k}$, i.e., \begin{equation}
 B_z\left(\mathbf{k}\right)=-\frac{k_x B_x\left(\mathbf{k}\right)+k_y B_y\left(\mathbf{k}\right)}{k_z}   
\end{equation} 
The zero frequency component (integration constant) was fixed by setting the average of $B_z$ to zero on the boundary of the reconstruction volume. To suppress noise amplification by this procedure, a butterworth-type low pass filter was applied.

\subsection*{Magnetic energy densities.}
Following Ref. \citep{Wolf2019} the exchange energy density may be split into contributions from magnetic charges, currents and surface terms $A\left( \left(\nabla\cdot\boldsymbol{M}\right)^{2}+\left|\nabla\times\boldsymbol{M}\right|^{2}-\mathit{w}_\mathrm{surf}\right)$. In the magnetostatic limit considered here, the magnetization in the second term may be replaced by $\boldsymbol{B}/\mu_0$ and can be reconstructed from the tomographic data. In the case of the DM interaction we have the following identities

\begin{align}
    E_{\text{DM}}\left[\boldsymbol{M}\right]
    &=D\int\boldsymbol{M}\cdot\left(\nabla\times\boldsymbol{M}\right)dV\\
	&=\frac{D}{\mu_{0}}\int\boldsymbol{B}\cdot\left(\nabla\times\boldsymbol{B}\right)dV+D\int\nabla\cdot\left(\Phi\boldsymbol{j}_{b}\right)dV\nonumber\\
	&=\frac{D}{\mu_{0}}\int\boldsymbol{B}\cdot\left(\nabla\times\boldsymbol{B}\right)dV+D\varoiint\boldsymbol{w}_\mathrm{surf}\cdot d\boldsymbol{S}\nonumber
\end{align}

Here $\boldsymbol{j}_b$ denotes the bound current and $\Phi$ the scalar magnetic potential. The last line identifies that part of the DM energy density, which can be derived solely from the $\boldsymbol{B}$-field, may be identified as a volume contribution, which can be reconstructed from tomographic data. The remainder can be collapsed to a surface term.

\subsubsection*{Acknowledgements}

We thank D. Pohl for helpful discussions in the process of planning the experimental setup and the data analysis. We furthermore acknowledge A. Tahn and T. Walter for the preparation of the FIB needle and magnetic ring, respectively. The authors are indebted to Vacuumschmelze GmbH \& Co.\ KG for providing the $\mathrm{Sm_2Co_{17}}$ magnet. AL, BR, and SS gratefully acknowledge financial support through the Priority Program SPP2137 of the German Research Foundation (DFG) within projects LU-2261/2-1 and RE-1164/6-1. DW and AL have received funding from the European Research Council (ERC) under the Horizon 2020 research and innovation program of the European Union (grant agreement number 715620). AK and RD received funding from the European Research Council (ERC) under the European Union's Horizon 2020 research and innovation programme (Grant No. 856538, project “3D MAGiC”),  from the European Union’s Horizon 2020 Research and Innovation Programme (Grant No. 823717, project “ESTEEM3”) and from the European Union’s Horizon 2020 Research and Innovation Programme (Grant No. 766970, project “Q-SORT”).

\subsection*{Author Contributions}

SS devised the experimental setup for the magnetic field stabilization. DW conducted the holographic VFET experiments with active support of SS and AK and performed the holographic and tomographic reconstructions. SS, DW, BR and AL analyzed the data. AL performed the magnetic simulations. AL, SS, DW, BR and UR wrote the manuscript. The FeGe single crystal was grown by MS. All authors contributed to the critical discussion and revision of the manuscript.

\subsection*{Supplementary Information}

Supplementary Information is available for this paper.

\subsection*{Author Information}

Correspondence and requests for materials should be addressed to
AL (a.lubk@ifw-dresden.de).

\bibliographystyle{naturemag}

\end{document}


\renewcommand{\thefigure}{S\arabic{figure}}
\setcounter{figure}{0}

\renewcommand{\thesection}{SI \arabic{section}}
\setcounter{section}{0}

\maketitle

\clearpage

\section{Specimen and tilt geometry}\label{suppl:note:geometry}
\begin{figure*}[ht]
    \centering
    \includegraphics[width=0.9\textwidth]{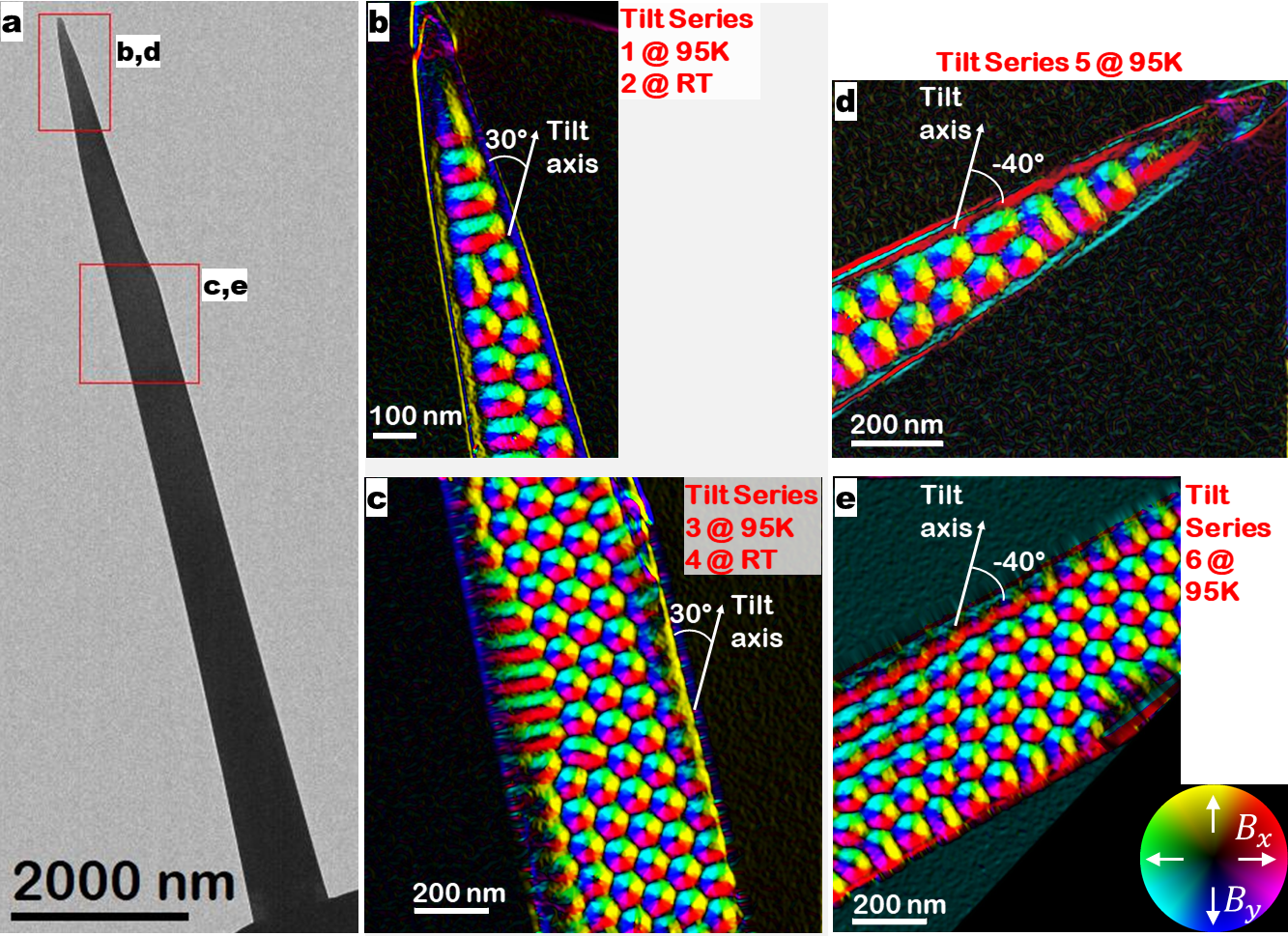}
    \caption{FeGe specimen and tilt geometry of the holographic VFET experiment. \textbf{a}, Bright-field TEM image of the ca.~10 µm long free-standing needle-shaped specimen. \textbf{b-e}, Reconstructed projected magnetic in-plane $\mathbf{B}$-field (color-wheel maps indicating the $x$- and $y$-direction of $\boldsymbol{B}$) at $0^\circ$ tilt angle and positions indicated in \textbf{a}.}
    \label{suppl:Fig_Specimen}
\end{figure*}

To perform the holographic VFET experiment at cryogenic conditions, we investigated a needle-shaped specimen prepared by focused ion beam (FIB) milling from a FeGe single crystal and transferred it to a lift-out TEM grid. Fig.~\ref{suppl:Fig_Specimen} shows the needle geometry and positions on the samples, where the tilt series were recorded (Fig.~\ref{suppl:Fig_Specimen}a). Moreover, Figs.~\ref{suppl:Fig_Specimen}b, c illustrate the orientation of the sample with respect to the tilt axis and the skyrmionic lattice arrangement for tilt series 1, 3, and Figs.~\ref{suppl:Fig_Specimen}d, e for tilt series 5, 6 after $70^\circ$ clock-wise in-plane rotation of the sample outside the TEM instrument at room temperature. Before rotation, two additional tilt series (2 and 4 in Figs.~\ref{suppl:Fig_Specimen}b, c) were acquired at room temperature (RT) and reconstructed in order to subtract their resulting electrostatic 3D phase maps from the ones obtained at $95\,\mathrm K$. Unfortunately, warming up the sample to room temperature, rotating the sample outside the microscope, and cooling down again to $95\,\mathrm K$ in the microscope led to a slight modification of the skyrmion lattice in the thicker region as clearly visible in Fig.~\ref{suppl:Fig_Specimen}c and Fig.~\ref{suppl:Fig_Specimen}e. Consequently, these data were too different for 3D magnetic vector field reconstruction.

\clearpage

\section{Magnetic ring}\label{suppl:note:ring}

\begin{figure}[ht]
    \centering
  \includegraphics[width=0.8\columnwidth]{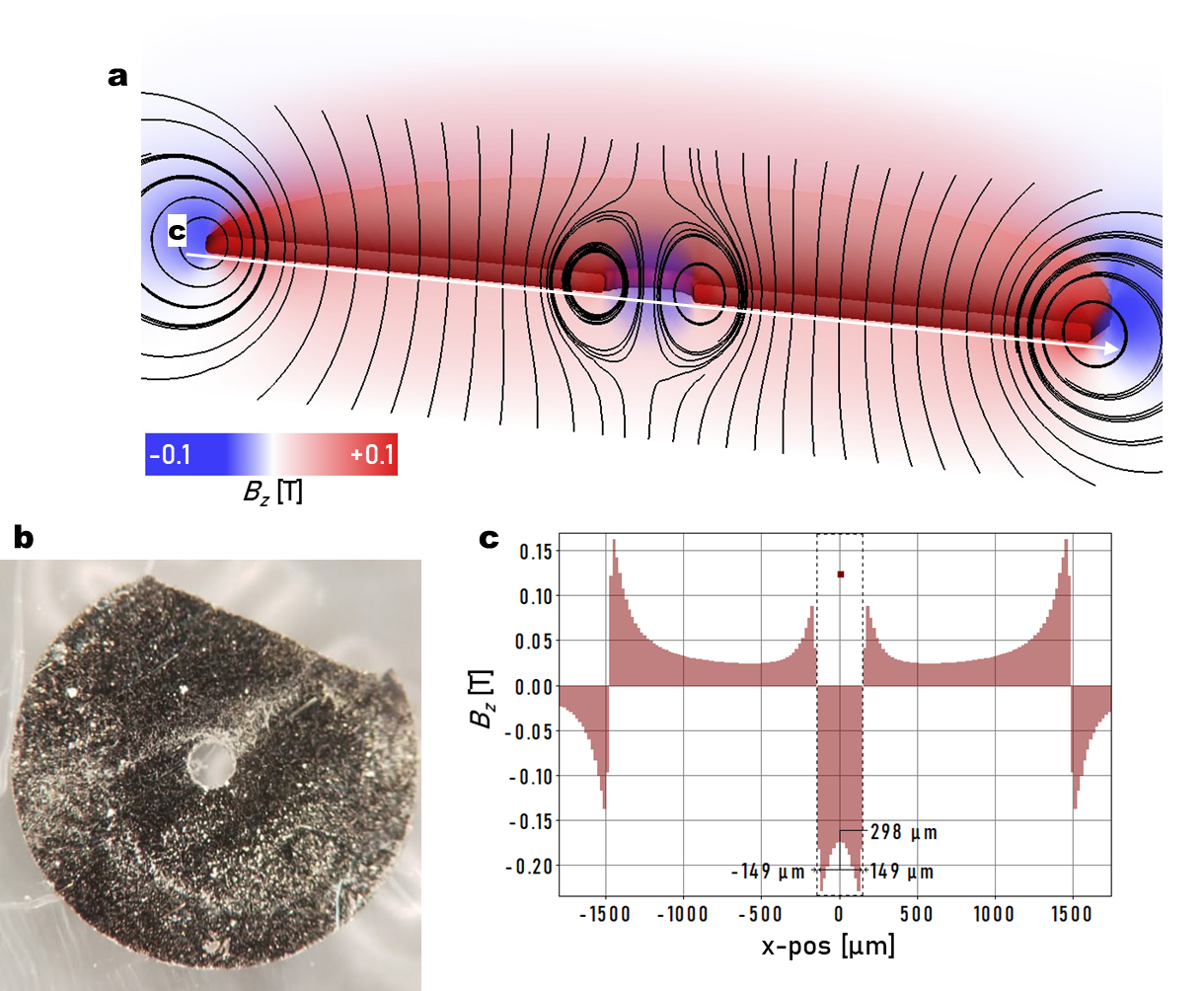}
  \caption{$\mathrm{Sm_2Co_{17}}$ permanent hard magnetic ring (50 \textmu m thickness with outer and inner diameter of 3 mm and 300 \textmu m) used for application of an external out-of-plane field on the FeGe specimen. \textbf{a}, Volume rendering (blue-white-red) of the $B_z$ component of the ring (cut through the center) obtained by magnetostatic simulations with black field-lines illustrating a $\boldsymbol{B}$-field slice through the center of the ring. \textbf{b}, Photograph of the ring after usage in the TEM. \textbf{c}, Line profile of the $B_z$-component taken at the lower surface of the ring at the position indicated by a white arrow in \textbf{a} (i.e., where the FeGe sample was placed in the experiment).}
  \label{suppl:Fig_MagRing}
\end{figure}

The skyrmion phase within the needle was stabilized by the stray field of a ring-shaped rare earth permanent magnet fitted into the liquid nitrogen cooling TEM holder. $\mathrm{Sm_2Co_{17}}$ was chosen for the ring material as it offers a high out-of-plane  saturation magnetization, high Curie point, good temperature stability and oxidation resistance \citep{SI:Gutfleisch2009}. The magnetic ring was prepared from a commercial cylindrical bulk magnet (outside radius: 1.5 mm, height: 1.5 mm) by sinker spark eroding and mechanical grinding to possess a hole with an inside radius of 150 \textmu m and a thickness of 50 \textmu m (cf. photograph in Fig. \ref{suppl:Fig_MagRing}b). The oxidation resistance is crucial during the preparation process, as only the surface of the $\mathrm{Sm_2Co_{17}}$ gets oxidized, whereby the magnetic performance of the ring is only slightly degraded \citep{SI:Chen2001}.

In order to test the properties of the ring after it's fabrication procedure, the magnetic induction was determined by magnetostatic simulations using the above-mentioned geometry and a value of the saturation magnetization of $1.1\, \mathrm{T}$ oriented in out-of-plane direction. The resulting $B_z$-component is shown in Fig.~\ref{suppl:Fig_MagRing}a. The red-blue contrast reveals that the stray-field within the hole points in the opposite direction of the magnetization as also illustrated by the black field lines. A line profile taken at the lower surface of the ring (Fig.~\ref{suppl:Fig_MagRing}c) confirms that the applied external field in out-of-plane direction is 170\, mT over a region of few tens of \textmu m in the center of the hole sufficient to stabilize the skyrmion phase in the FeGe specimen.


\clearpage

\section{3D electrostatic mean inner potential (MIP)}\label{suppl:note:MIP}

\begin{figure*}[ht]
    \centering
    \includegraphics[width=0.9\textwidth]{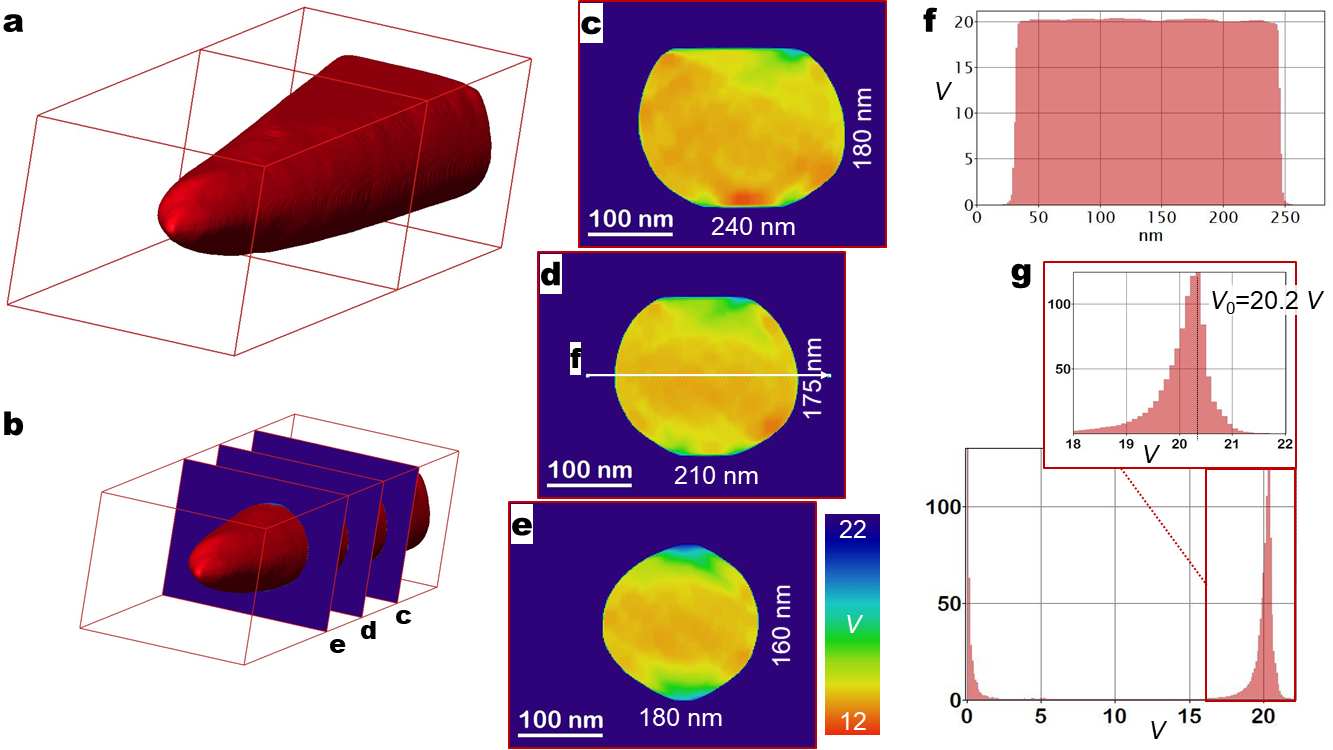}
    \caption{3D electrostatic potential of the FeGe needle-shaped FIB-prepared sample reconstructed by electron holographic tomography at room temperature. \textbf{a}, 3D Iso-potential surface ($15\,\mathrm{V}$) visualizing the overall morphology. \textbf{b}, Same as a but with slices included to display three cross-sections, \textbf{c-e}, from bottom to the top of the sample. \textbf{f}, Potential profile along the line scan indicated as arrow in \textbf{d} revealing a plateau of $20\,\mathrm{V}$ that is consistent with a peak in the histogram \textbf{g} of the FeGe sample at $20.2\,\mathrm{V}$.}
    \label{suppl:Fig_MIP}
\end{figure*}

A 3D reconstruction of the FeGe needle-shaped FIB-prepared sample at room temperature was carried out, in order to obtain the electrostatic contribution to the 3D phase shift

\begin{align}
    \dfrac{\partial\varphi_{el}}{\partial z}\left(x,y,z\right)=C_\text{E} V_\text{0}\left(x,y,z\right)
\end{align}

that is proportional (interaction constant for 300 kV electrons is $C_\mathrm{E}=0.0065\,\mathrm{rad}/\mathrm{Vnm}$) to the 3D electrostatic potential, i.e.,  mainly the mean inner potential (MIP). The MIP is of particular interest for quantitative 2D electron holography, to determine the thickness of the sample\citep{SI:Schneider2018}. As described in the methods section of the main text, the tilt series were recorded from $-66^\circ$ to $+65^\circ$, hence the reconstructed tomogram suffers from missing wedge artefacts due to the incomplete tilt range ($\pm{90}^\circ$). These missing wedge artefacts in Fourier space reduce the resolution in the directions not covered in the projection data and hamper an accurate measurement of the MIP. For this reason, we applied a so-called finite support approach for missing wedge correction at low spatial frequencies in the tomogram\citep{SI:Wolf2018}, prior the 3D analysis of the potential data. Fig~\ref{suppl:Fig_MIP} summarizes the 3D electrostatic potential reconstruction.

\clearpage

\section{Spatial resolution of tomographic reconstruction}\label{suppl:note:resolution}

\begin{figure*}[ht]
    \centering
    \includegraphics[width=0.9\textwidth]{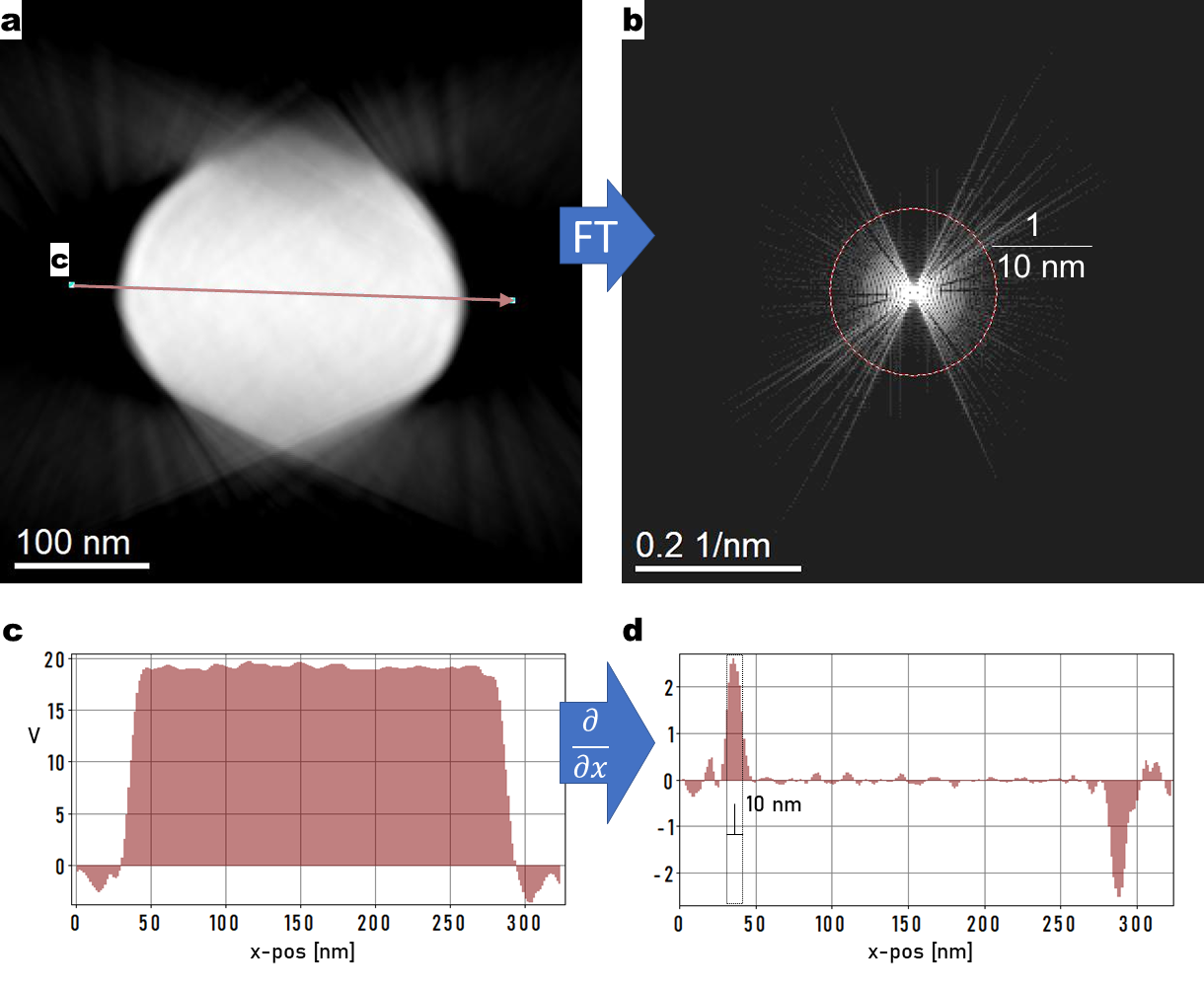}
    \caption{Spatial Resolution of tomographic reconstruction. \textbf{a}, Representative slice of the MIP tomogram showing the cross-section of the FeGe needle-shaped FIB sample. \textbf{b}, Fourier transform of \textbf{a}. \textbf{c}, Line profile along the line scan in indicated be an arrow in \textbf{a}. \textbf{d}, Numerical derivative of \textbf{c}.}
    \label{suppl:fig:resolution}
\end{figure*}

VFET involves a complex reconstruction process and data treatment from the raw data (tilt series) band-limited in both projection space (detector sampling) and angular space (finite tilt steps, limited tilt range) to the final tomograms. This data treatment includes, e.g., interpolation, Fourier and spatial filtering, as well as regularization within the tomographic reconstruction algorithm. Therefore, it is not possible to describe the signal and noise propagation in a straightforward manner, and to provide an overall isotropic lateral resolution value for the obtained tomograms. Here, we therefore provide an experimental resolution measure obtained from the sharpness of the object edge in the MIP tomogram (Fig. \ref{suppl:fig:resolution}). Fig. \ref{suppl:fig:resolution}a shows an $x$-$z$ slice of the reconstructed MIP tomogram of the FeGe FIB needle, Fig. \ref{suppl:fig:resolution}b its Fourier transform (FT). In the latter, the superposition of central slices (the FT of the projections) is clearly visible as well as the missing wedge in vertical direction beyond the experimental tilt range. The red dashed circle illustrates the band, in which spatial frequencies until $0.1\, \mathrm{nm^{-1}}$ are transferred corresponding to a $10\,\mathrm{nm}$ resolution in real space. A line profile is depicted in Fig. \ref{suppl:fig:resolution}c along the line scan through the cross-section in Fig. \ref{suppl:fig:resolution}a to obtain the edge spread functions on both sides of the object. Its width is determined by computing the derivative in direction of the line scan, e.g. in $x$-direction, yielding the so-called line spread function. We assign the full width half maxima (FWHM) of the two peaks of $10\,\mathrm{nm}$ to an estimated value of the lateral resolution. In Ref. \citep{SI:Wolf2019} we used Fourier shell correlation to assess the lateral resolution of similar reconstructed VFET data in a more elaborated way, obtaining a value of $7\,\mathrm{nm}$. 

\clearpage

\section{Skyrmion lattice packing}\label{suppl:note:Lattice Packing}

\begin{figure*}[ht]
  \centering
  \includegraphics[width=1.0\textwidth]{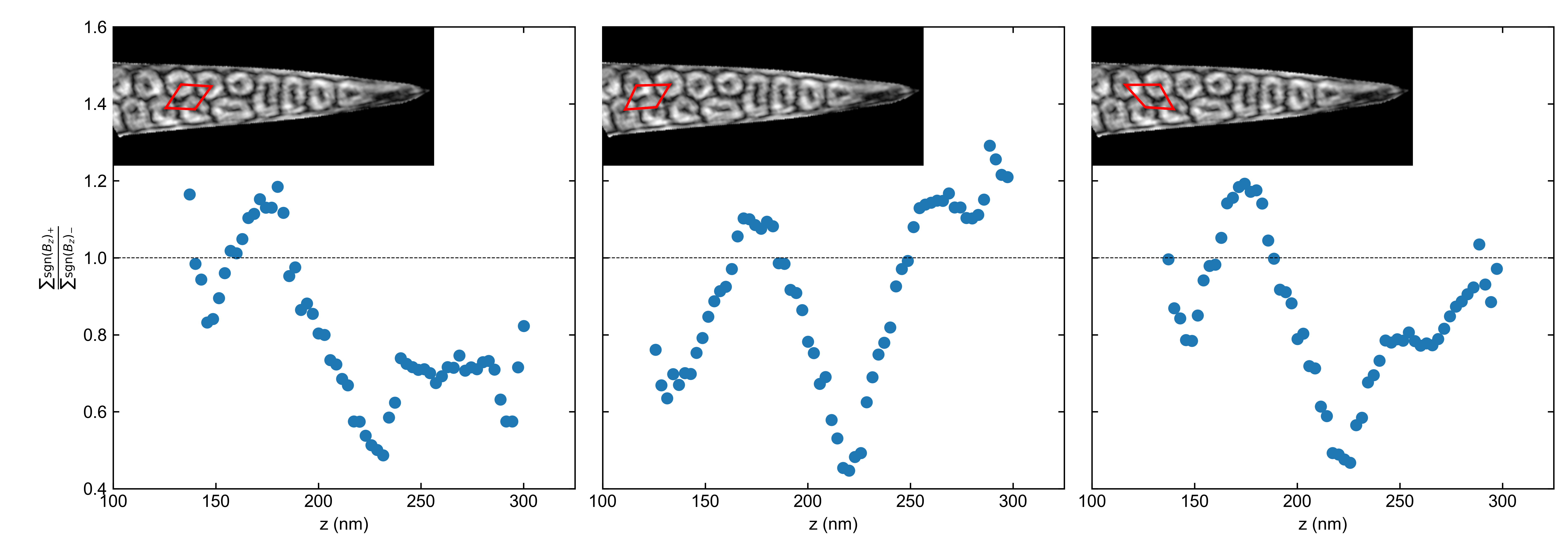}
  \caption{Proportion of positive and negative $B_z$ values inside the magnetic unit cells (red rhombi in the insets) as function of the $z$ position in the 3D stack.}
  \label{Signum Bz}
\end{figure*}

In order to analyze the packing of the skyrmions within the needle, the ratio of positive and negative $B_z$ values inside a magnetic unit cell was calculated. To do so the centres of the skyrmions for each $z$ slice of the 3D stack were determined from the in-plane magnetic conduction $B_{\perp} = \sqrt{B_x^2 + B_y^2}$. Connecting these points yields three magnetic unit cells within the field view (cf. red rhombi in the insets in Fig.~\ref{Signum Bz}). These unit cells were subsequently utilized to calculate the proportion of the areas, where $B_z$ is pointing upward and downward. The resulting values for each $z$ slice of the three unit cells are plotted in Fig.~\ref{Signum Bz}. The graphs show a non-trivial behaviour as function of the $z$ position, which can be partly attributed to reconstruction artifacts of $B_z$. Nevertheless all three graphs manifest a ratio below 1 for most of the slices. Taking into account the limited number of samplings a mean value for the ratio of 0.85 was determined, indicating a closed packed skyrmion lattice \citep{SI:Balkind2019}.

\clearpage

\section{Magnetic energy densities}\label{suppl:note:mag-energies}

\begin{figure*}[!ht]
  \centering
  \includegraphics[width=0.75\textwidth]{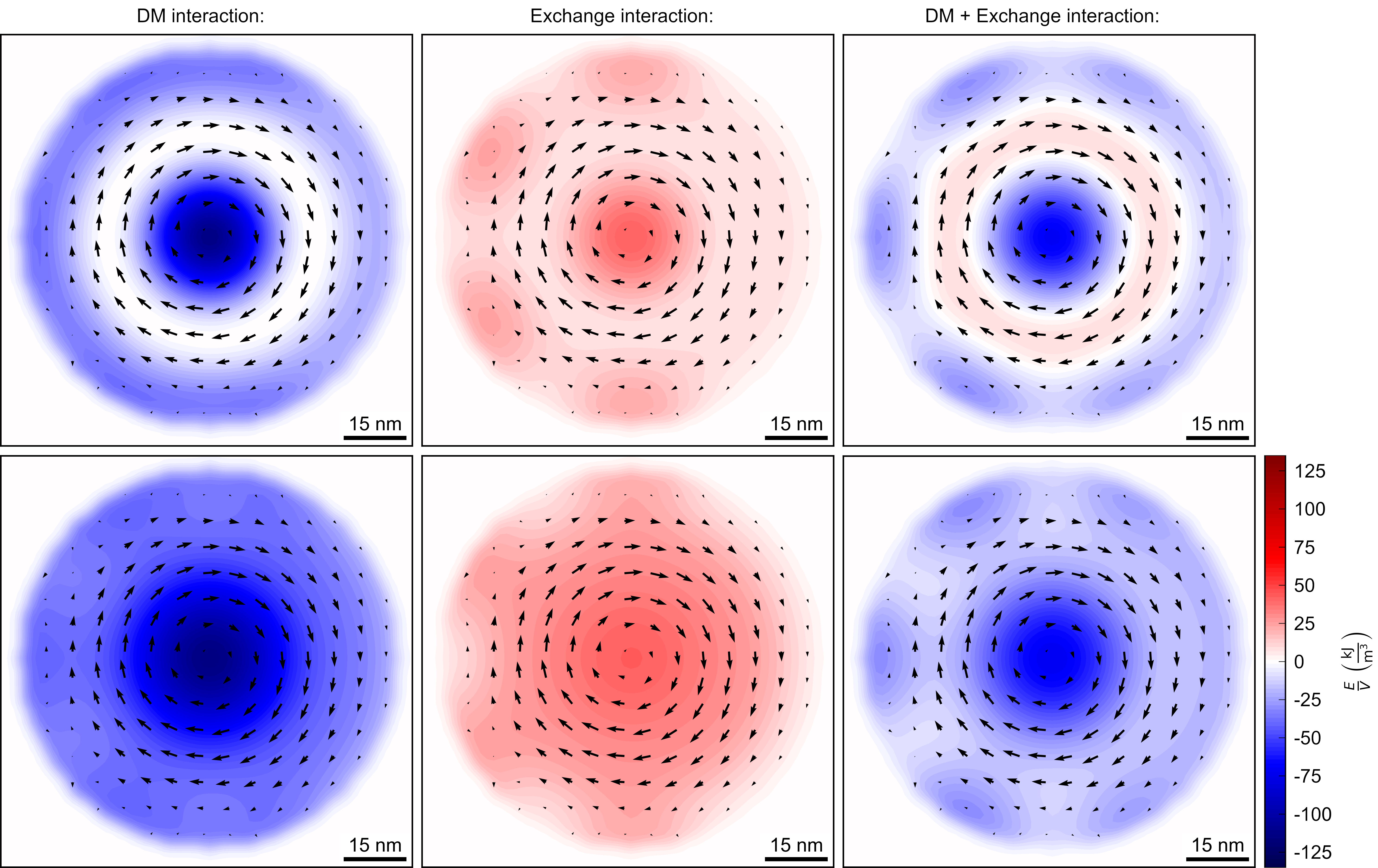}
  \caption{Representative magnetic energy density contributions $A\|\left(\nabla \times \boldsymbol{B}\right)_z \|^2$ (exchange interaction) and $D \boldsymbol{B}_z \cdot ( \nabla \times \boldsymbol{B})_z$ (DM interaction) from magnetostatic simulations (upper row). Corresponding energy contributions $A\|\left(\nabla \times \boldsymbol{B}\right) \|^2$ and $D \boldsymbol{B} \cdot ( \nabla \times \boldsymbol{B})$ (lower row). The arrows indicate the in-plane magnetic induction.}
  \label{Suppl_Fig_energetics_simulation}
\end{figure*}

As already described in the main text, the representations of the energy densities (cf. Fig. \ref{main:Fig_4} in the main text) were limited to the contributions dominated by the in-plane components of the magnetic induction due to persisting reconstruction artifacts in the $B_z$ component. This restriction, however, is sufficient to reveal the skyrmion energetics. The comparison of the energy terms calculated with the in-plane and total magnetic induction from magnetostatic simulations (cf. Fig.~\ref{Suppl_Fig_energetics_simulation}) reveals, that the energetics of the skyrmion core are dominated by $B_z$. In contrast the adjacent ring of in-plane magnetization is mainly determined terms containing $B_x$ and $B_y$.      

The described restriction is consistently applicable to all skyrmions as displayed in Fig. \ref{Suppl_Fig_energetics_simulation}. In most of the skyrmion cores there is a lowering of the energy sum observable. 

\begin{figure*}[!ht]
  \centering
  \includegraphics[width=0.9\textwidth]{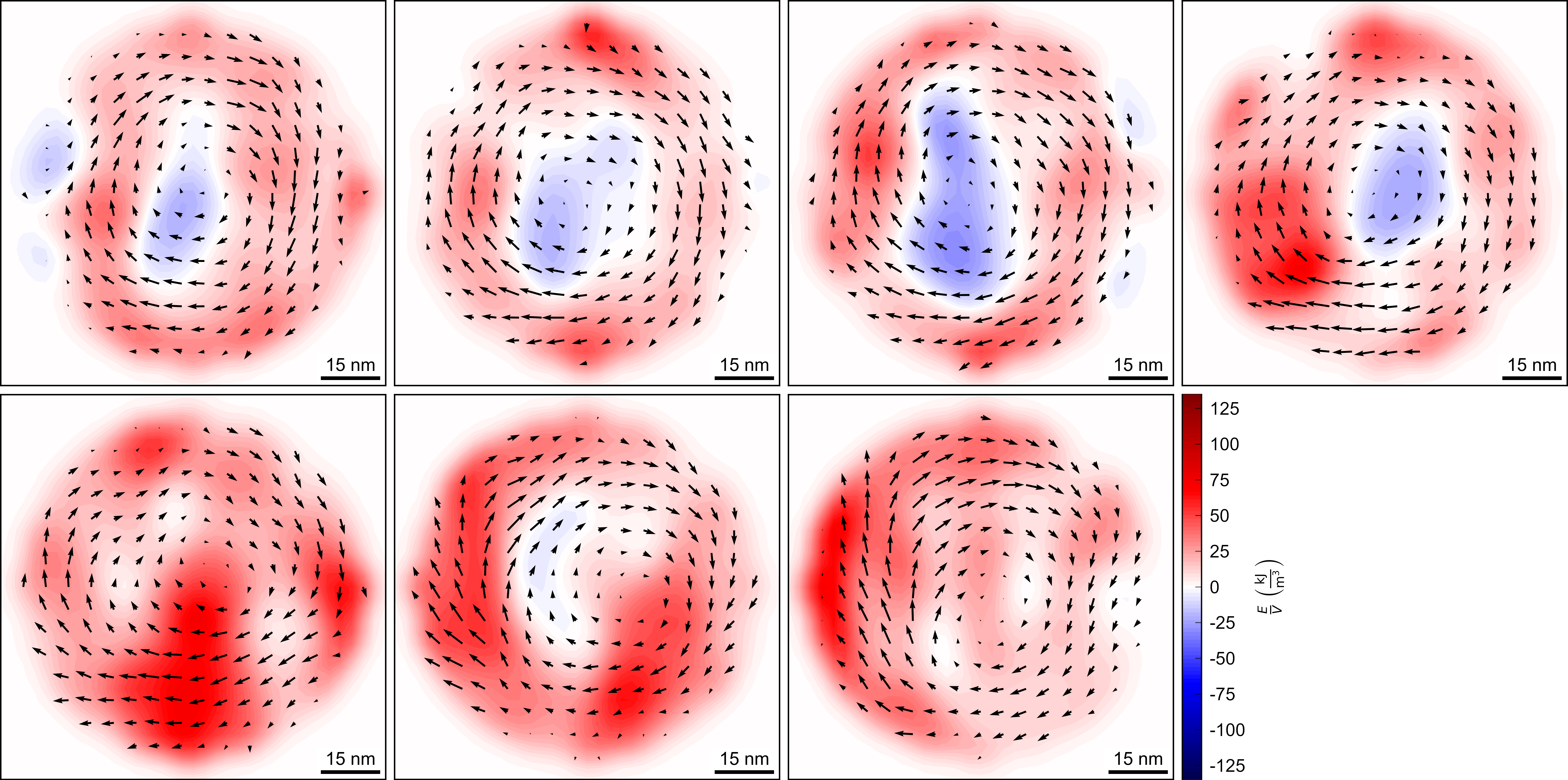}
  \caption{Sum of DM and exchange energy densities dominated by the experimental magnetic in-plane induction for all skyrmions. The arrows indicate the in-plane magnetic induction.}
  \label{energetics_experiment}
\end{figure*}

\clearpage

\section{Simulation of magnetic fields pertaining to skyrmion lattice in the needle}\label{suppl:note:Magnetostatics}

To generate a theoretical magnetic model of the skyrmion lattice in the needle required for energy density considerations (main text and \ref{suppl:note:mag-energies}) and the computation of artificial tomographic tilt series we carried out the following steps: (I) Synthesis of an artificial magnetic skyrmion texture by assembling a double line of SkTs with the help of the Circular Cell Approximation \citep{Bogdanov1994suppl}. Herein we used the exchange constants $A = 8.78 \ \mathrm{\frac{pJ}{m}}$, $D = 1.58 \ \mathrm{\frac{mJ}{m^2}}$  \citep{Beg2015suppl}, an experimentally determined circular cell radius of 40 nm, and an external field of 170 mT as produced by the magnetic ring (Sec. \ref{suppl:note:ring}). Dipolar interaction was taken into account through effective easy-plane anisotropy approximation in thin films \citep{Bogdanov1994suppl}.  The numerical solution of the non-linear differential equation as a boundary value problem pertaining to the polar angle of the magnetization in the circular cell was carried out with the help of the shooting method implemented in the DifferentialEquations package of Julia programming language. (II) In between the circular cells magnetization pointing into $z$-direction along the external fields was assumed. Truncation to the outer shape of the nanowire was imposed by multiplication with a shape function of the needle, (III) Solution of the magnetostatic equations (2)-(4)
\begin{align}
    \Delta \Phi_M=\mathrm{div}\,\mathbf{M}
\end{align}

\begin{align}
    \mathbf{H}=-\nabla\,\Phi_M
\end{align}

\begin{align}
    \mathbf{B}=\mu_0\left(\mathbf{M}+\mathbf{H}\right)
\end{align}
within a supercell containing vacuum around the needle imposing periodic boundary conditions. The supercell (i.e., vacuum region) was chosen large enough to suppress boundary effects. (IV) To compute the energy density maps we additionally convoluted the obtained magnetic fields with a Gaussian smoothing kernel corresponding to the experimental spatial resolution.

\clearpage

\section{Surface modification }\label{suppl:note:surface layer}

Unlike homogeneous bulk materials, the investigated needle is nanoscopically small with a large surface-area-to-volume ratio. As a consequence, the magnetic texture can be easily modified in these surface regions influencing the structure of the whole skyrmion. The effect of the sample preparation by focussed ion beam (FIB) milling was investigated, since the highly energetic ions are partially implanted in the surface area of the needle, and may create defect cascades and even lead to partial amorphization. Hence, to understand the effect of FIB-structuring, the defect landscape of the cross-section of an equivalent FeGe FIB lamella was characterised with HRTEM and EDX mapping. After preparation of a conventional FIB FeGe sample, the lamella was coated with amorphous carbon. Subsequently a thin slap was cut out of the sample to investigate the cross-section of the lamella.

\begin{figure}[ht]
    \centering
  \includegraphics[width=0.75\columnwidth]{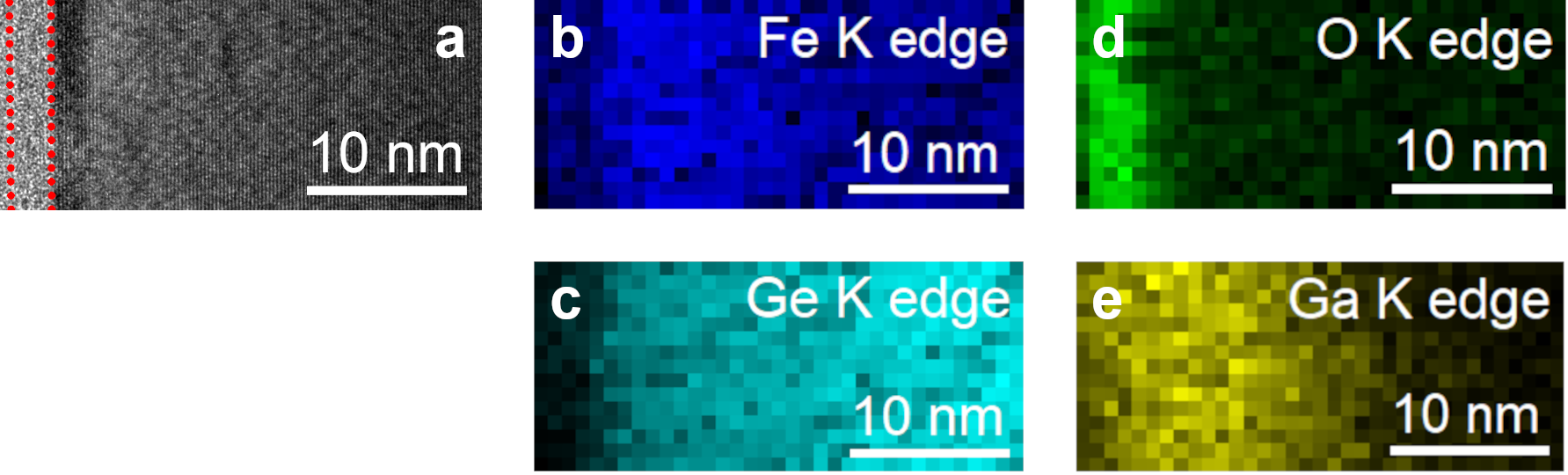}
  \caption{HRTEM and STEM-EDX investigations of the crystalline structure and elemental composition of the surface of a FeGe lamella prepared by FIB. \textbf{a}, The HRTEM measurement show an amorphous layer (marked by red dotted lines) on top of the crystalline core. \textbf{b-e}, Complementary STEM-EDX investigations reveal, that this amorphous layer is an iron oxide, which results from the removal of Ge by the $\mathrm{Ga^{+}}$ ions during FIB procedure.}
  \label{suppl:Fig_HRTEM_EDX}
\end{figure}

HRTEM measurements (cf. Fig. \ref{suppl:Fig_HRTEM_EDX}a) reveal that the inner part of the sample is still fully crystalline. This core, however, is surrounded by a roughly 4 nm thin shell, which is marked by red dotted lines. Complementary STEM-EDX investigations show that there is still iron present in this amorphous layer (cf. Fig. \ref{suppl:Fig_HRTEM_EDX}b), but no germanium (cf. Fig. \ref{suppl:Fig_HRTEM_EDX}c). Instead a strong oxygen signal can be found at the surface (cf. Fig. \ref{suppl:Fig_HRTEM_EDX}d). Furthermore a strong gallium signal can be found, which reaches up to 17 nm into the lamella volume in this sample area (cf. Fig. \ref{suppl:Fig_HRTEM_EDX}e). Thus we conclude that $\mathrm{Ga^{+}}$ ions are implanted into the lamella during the FIB milling and preferentially removed germanium from the sample surface. The unbound iron gets oxidized and forms an amorphous layer.

Although the investigated FeGe needle most likely also features such an amorphous surface, these layers can only partly account for the complicated magnetic texture of the skyrmions, which is disturbed almost throughout the whole volume of the needle, where the sample is still perfectly intact.

\clearpage

\section{Hologram centerband and sidebands}\label{suppl:note:solitonic skyrmions}
\begin{figure}[ht]
  \centering
  \includegraphics[width=0.9\columnwidth]{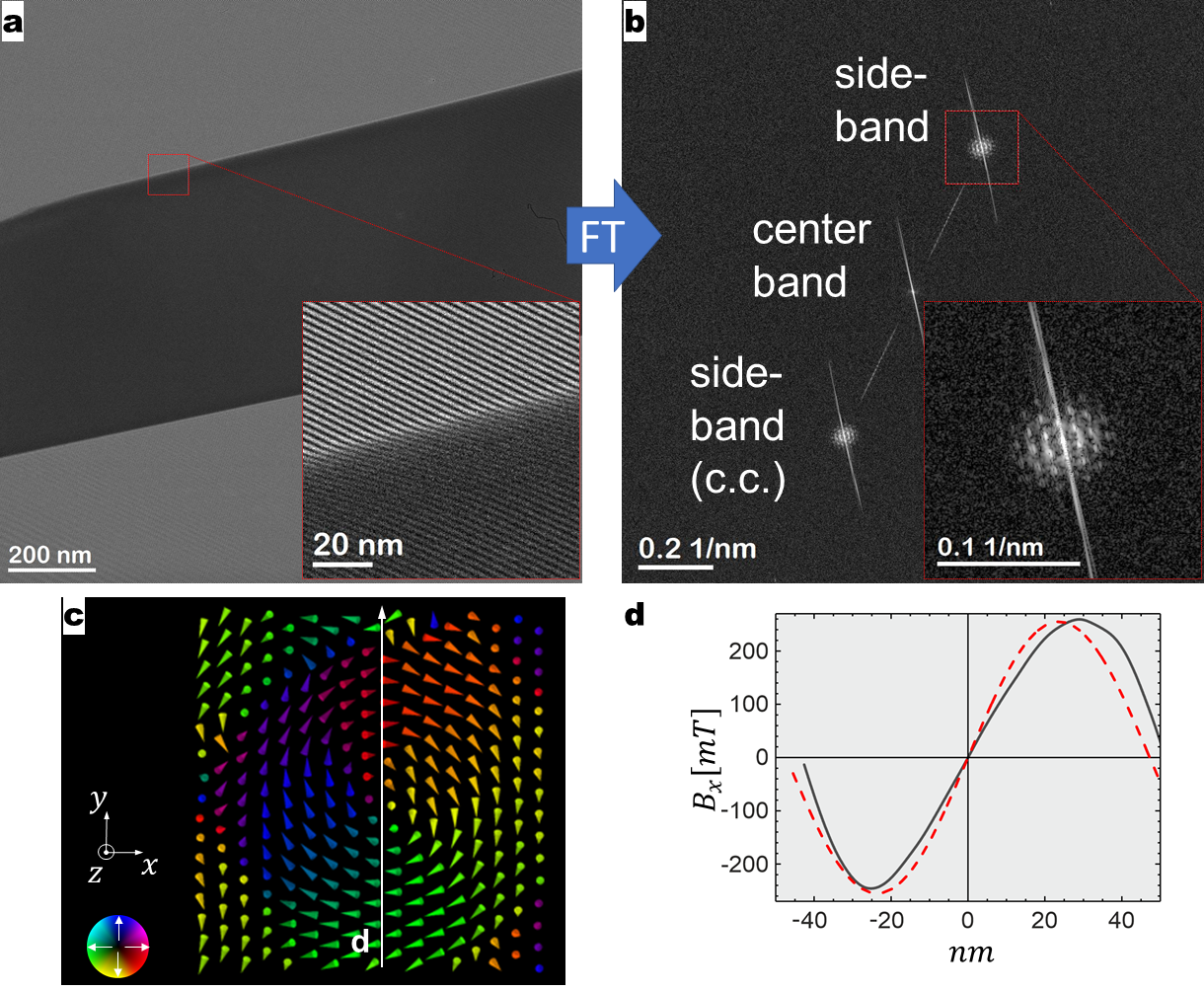}
  \caption{Observation of solitonic nature of skyrmions. \textbf{a}, Electron hologram of the FeGe needle-shaped FIB-prepared sample taken at $95\,\mathrm K$ at position shown in  Fig.~\ref{suppl:Fig_Specimen}c. \textbf{b}, Fourier transform of \textbf{a} showing higher order reflections of the skyrmion lattice in holographic sideband (inset). \textbf{c}, Arrow plot of an in-plane ($x$-$y$) slice through a skyrmion reconstructed by VFET. \textbf{d}, Line profile of the $B_x$ component taken along the line scan indicated by the white arrow in \textbf{c}.}
  \label{suppl:Fig-sideband}
\end{figure}

Analyzing the Fourier transform of the electron wave, i.e., the holographic sideband visible in the Fourier transform of the electron hologram (cf. Fig.~\ref{suppl:Fig-sideband}a) recorded in the thicker FeGe specimen region, we observe up to $\mathrm{3^{rd}}$ order systematic reflections of the magnetic skyrmion lattice (cf. Fig.~\ref{suppl:Fig-sideband}b). They demonstrate the solitonic nature of the skyrmions, i.e., the arrangement of solitonic field modulations with a non-trivial structure factor in an ordered lattice. Alternatively, they may be also observed by investigating the rather pronounced non-harmonic rotation of the field when striding away from the center of a single skyrmion tube in the reconstructed 3D field. Fig.~\ref{suppl:Fig-sideband}c shows an in-plane ($x$-$y$) slice through a skyrmion reconstructed by VFET (cf. Fig.~\ref{main:Fig_2} in the main text). The arrow (parallel to the $y$-direction) indicates, where the line profile of the $B_x$ component Fig.~\ref{suppl:Fig-sideband}d is taken. It shows a larger slope of the modulation close to the skyrmion core that agrees well to theoretical models of single skyrmions and highly resolved experimental measurements with spin-polarized STM \citep{SI:Leonov2016}.

\clearpage

\section{Electric and magnetic fringing fields}\label{suppl:note:el_mag_fringing_fields}
\begin{figure*}[ht]
  \centering
  \includegraphics[width=0.9\textwidth]{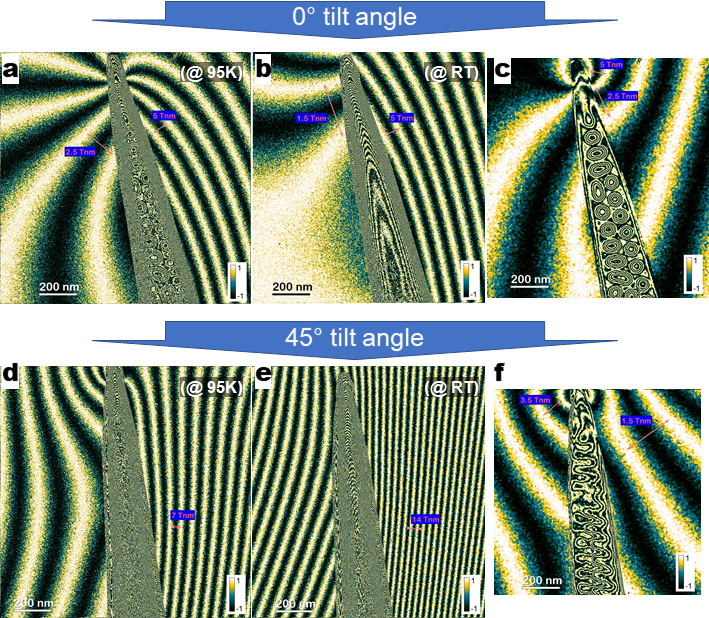}
  \caption{Fringing fields around the FeGe specimen observed in phase images. \textbf{a, b}, Cosine of phase images ($10\times$ amplified) illustrating the projected in-plane $\mathbf{B}$-field components taken at $0^\circ$ and \textbf{a} $95\, K$, room temperature \textbf{b}. \textbf{c}, Magnetic phase image obtained by subtraction of \textbf{a} and \textbf{b}. \textbf{d-f}, Same as \textbf{a-c} but at $45^\circ$ tilt angle.}
  \label{Suppl_Fig-fringing fields}
\end{figure*}

The FeGe needle specimen as well as the magnetic ring cause magneto-static fringing fields. In case the sample is slightly charged up by the electron beam, also electro-static stray fields may occur. If these far-reaching fields are too strong, they may disturb the tomographic reconstruction. Here, we demonstrate that these fields are small enough to do not influence significantly the 3D reconstruction. Fringing fields are visible in the vacuum regions of the phase images as phase modulation. Fig.~\ref{Suppl_Fig-fringing fields} shows the cosine of $10\times$ amplified phase images providing fringes that illustrate direction and strength of the projected in-plane components of the magnetic induction. In Fig.~\ref{Suppl_Fig-fringing fields}a, b, phase images at $0^\circ$ tilt angle display similar density of the projected field lines in the vacuum for 95K (skyrmion state) and room temperature (non-magnetic), hence the major contribution of the stray field comes from the magnetic ring. Consequently, after subtraction of both phase images and correction of a constant phase gradient, in the resulting magnetic phase image (cf. Fig.~\ref{Suppl_Fig-fringing fields}c) a smaller stray field is left. The same is true for phase images reconstructed at $45^\circ$ (cf. Fig.~\ref{Suppl_Fig-fringing fields}d-f. Quantitative evaluation of the fringe distances in the magnetic phase images result in projected fields between $1.5\,$Tnm and $3.5\,$Tnm depending on tilt angle and position. Considering a back-projection path length of ca. $300\,$nm used for tomographic back-projection of the projection data, this corresponds to $\boldsymbol{B}$-field variations in the tomogram in the range of $0.01\,$T.

\clearpage

\section{Tomographic reconstruction of simulated skyrmion needle}\label{suppl:note:Rec-of-Sim}

\begin{figure*}[!ht]
    \centering
    \includegraphics[width=0.8\textwidth]{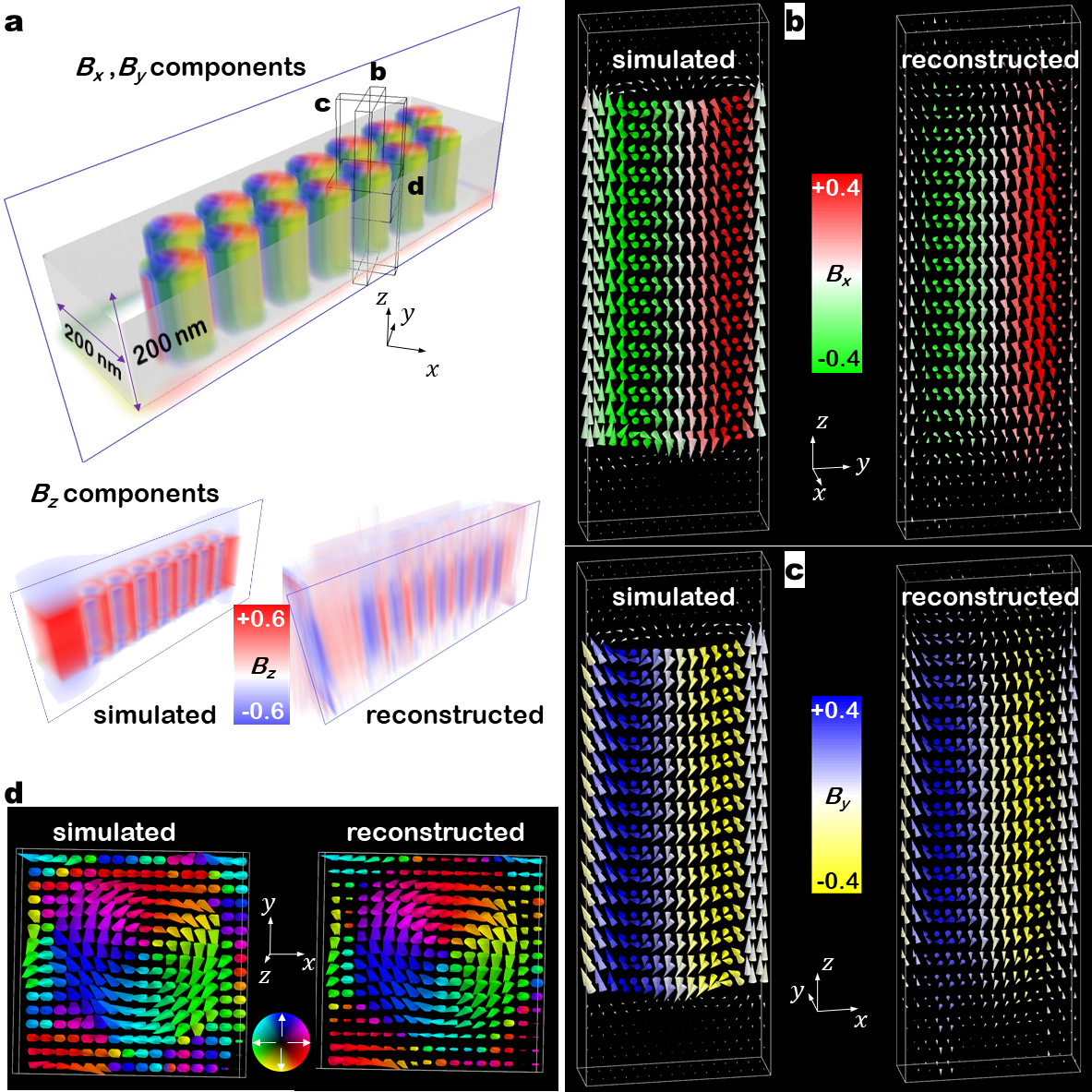}
    \caption{Comparison of simulated skyrmion needle with its tomographic reconstruction using the same reconstruction workflow and parameters as in the experimental case. \textbf{a}, Color-coded volume rendering of simulated $B_{x}$ (green-red) and $B_y$ (yellow-blue) components. Likewise to the experimental case (see \textbf{Fig. \ref{suppl:Fig_Specimen}}), the needle-axis is rotated $30^\circ$ to the $x$ axis. The simulated and reconstructed $B_z$ components are volume-rendered in blue-red. The elongation (blurring) of the $B_z$ reconstruction along $z$ direction due to the incomplete tilt range is clearly visible. \textbf{b}, According to \textbf{Fig. \ref{main:Fig_2}} of the main text, 2D cross sections along \textbf{b} the $y$-$z$, \textbf{c}, the $x$-$z$, and \textbf{d}, the $x$-$y$ planes as indicated in \textbf{a} are taken for both cases.} 
    \label{suppl:Fig_rec-sim}
\end{figure*}

In order to verify the fidelity of $\boldsymbol{B}$-field of the FeGe needle reconstructed in 3D from the experimental tilt series, we mimicked a tomographic vector-field reconstruction of a simulated magnetic skyrmion needle (cf. Fig.~\ref{suppl:Fig_rec-sim}a) with similar dimensions, orientation, and sampling as in the experimental case (see Methods section of main text). In detail, we employed the following procedure:

\begin{enumerate}
    \item Magnetostatic simulation of the entire 3D magnetic induction of the needle (see Sect.~\ref{suppl:note:Magnetostatics}) as shown in Fig. \ref{suppl:Fig_rec-sim}a. The dimensions of the needle in $x$-, $y$-, and $z$- direction are $200\,\mathrm{nm}$ by $750\, \mathrm{nm}$ by $200\, \mathrm{nm}$. The orientation of the skyrmion tubes are along $z$-direction.
    \item Rotation of the simulated needle by $30^\circ$ around $z$-axis as in the experimental case.
    \item Computation of the magnetostatic vector potential $\boldsymbol{A}$ by solving $-\Delta \boldsymbol{A}=\nabla\times\boldsymbol{B}$ (Coulomb gauge).
    \item Simulation of the electrostatic scalar potential $V(x,y,z)$ using a constant mean inner potential (MIP) of $22\, \mathrm{V}$ within the needle.
    \item Computation of the phase tilt series tilted about angles $\alpha$ around the $y$-axis using the experimental tilt range. For each $y$-position, the line integrals along $\mathrm{d}\mathbf{s}$ depending on $\alpha$ and $l$ the coordinate in the projection line read for the electrostatic part 
    \begin{equation}
        \varphi_{el}\left(l,y,\alpha\right)=C_{E}\int_{-\infty}^{+\infty} V\left(x,y,z\right)\mathrm{d}\mathbf{s}
    \end{equation}
    and for the magnetostatic part
    \begin{equation}
        \varphi_{mag}\left(l,y,\alpha\right)=\\-\frac{e}{\hbar}\int_{-\infty}^{+\infty}\left( A_x\left(x,y,z\right)\mathrm{sin(\alpha)}+A_z\left(x,y,z\right)\mathrm{cos(\alpha)}\right)\mathrm{d}\mathbf{s}
    \end{equation}
    \item Rotation of the simulated needle by -$60^\circ$ around $z$-axis and repetition of steps 3-5 to compute the second phase tilt series with the same parameters as in the experiment.
    \item Calculation of phase tilt series $\varphi=\varphi_{el}+\varphi_{mag}$ 
    \item Tomographic reconstruction of three 3D phase maps, two total ones rotated by 90° to each other, and an electrostatic one from the corresponding tilt series
    \item Subtraction of the electrostatic 3D phase map from the other two 3D phase maps to obtain the magnetic 3D maps as described in the Methods part of the main text
    \item Derivation of the two magnetic 3D phase maps in direction parallel to their tilt axes and multiplication with the factor $\hbar /e$ leading to the corresponding $\boldsymbol{B}$-field components.
    \item Calculation of $B_{z}$ from $B_{x}$ and $B_{y}$ as described in the Methods part of the main text.
\end{enumerate}

The results are depicted in Fig.~\ref{suppl:Fig_rec-sim} visualized in a similar way as Fig.~\ref{main:Fig_2} of the main text. It clearly shows that all three $\boldsymbol{B}$-field components can be reconstructed in a remarkable quality using the here presented workflow applied to a tilt series with incomplete tilt range. The limitations of the tomograms are mainly noticeable by a reduced resolution in $z$-direction (missing wedge artefacts), in particular visible in the $B_z$-component (cf. Fig.~\ref{suppl:Fig_rec-sim}a), but also in the other two components the elongation and gradual reduction of the signal along $z$-direction (cf. Fig.~\ref{suppl:Fig_rec-sim}b, c) is visible.

\clearpage

\bibliographystyle{naturemag}